\documentclass[aps,superscriptaddress,amsmath,amssymb,twocolumn,showpacs,floatfix,english,noshowpacs]{revtex4}

\usepackage{url}
\usepackage{bm,bbm}
\usepackage{graphicx}
\usepackage{color}
\usepackage[colorlinks=true, urlcolor=blue, linkcolor=blue, citecolor=blue, pdftex]{hyperref}
\usepackage[english]{babel}

\newcommand{\dagga}{{\phantom{\dagger}}}

\begin{document}

\title{Dynamical structure factor of the $J_1-J_2$ Heisenberg model in one dimension: The variational Monte Carlo approach}

\author{Francesco Ferrari}
\email[]{frferra@sissa.it}
\affiliation{SISSA-International School for Advanced Studies, Via Bonomea 265, I-34136 Trieste, Italy}
\author{Alberto Parola}
\affiliation{Dipartimento di Scienza e Alta Tecnologia, Universit\`a dell'Insubria, Via Valleggio 11, I-22100 Como, Italy}
\author{Sandro Sorella}
\affiliation{SISSA-International School for Advanced Studies, Via Bonomea 265, I-34136 Trieste, Italy}
\author{Federico Becca}
\affiliation{Democritos National Simulation Center, Istituto Officina dei Materiali del CNR and 
SISSA-International School for Advanced Studies, Via Bonomea 265, I-34136 Trieste, Italy}

\date{\today}

\begin{abstract}
The dynamical spin structure factor is computed within a variational framework to study the one-dimensional $J_1-J_2$ 
Heisenberg model. Starting from Gutzwiller-projected fermionic wave functions, the low-energy spectrum is constructed 
from two-spinon excitations. The direct comparison with Lanczos calculations on small clusters demonstrates the excellent 
description of both gapless and gapped (dimerized) phases, including incommensurate structures for $J_2/J_1>0.5$. 
Calculations on large clusters show how the intensity evolves when increasing the frustrating ratio and give an 
unprecedented accurate characterization of the dynamical properties of (non integrable) frustrated spin models.
\end{abstract}

\maketitle

\section{Introduction}\label{sec:intro}

Quantum spin liquids are unconventional phases of matter in which quantum fluctuations endure any tendency to develop 
local (e.g., magnetic) order, down to zero temperature. Most importantly, the lack of any symmetry breaking mechanism is 
accompanied by topological order and the presence of excitations with fractional quantum numbers~\cite{balents2010}. For
concreteness, let us consider the spin-$1/2$ Heisenberg model on a given two-dimensional lattice. Whenever magnetic order 
sets in (e.g., on the square lattice with nearest-neighbor antiferromagnetic interactions), the elementary excitations 
are $S=1$ magnons or spin waves, which can be pictured as coherent Bloch waves made from localized spin flip excitations. 
Instead, in spin liquids, the elementary objects are $S=1/2$ spinons, while $S=1$ excitations decay in two spinons that 
are asymptotically free at long distances~\cite{balents2010}. Since, for any lattice with a fixed number of sites, the 
minimal change in the total spin is $\Delta S= \pm 1$, the existence of objects with $S=1/2$ implies a fractionalization 
of the spin quantum number. Spinons do exist in the one-dimensional Heisenberg model~\cite{faddeev1981}, where the magnetic 
long-range order is hampered in agreement with the generalized Mermin-Wagner theorem~\cite{pitaevskii1991}. In this case, 
the spectrum is gapless and characterized by the presence of a broad continuum of excitations. A similar feature is present 
in the one-dimensional Heisenberg model with inverse-square superexchange~\cite{haldane1988}; in this case, spinons are non 
interacting, and the whole excitation spectrum can be found explicitly in a closed form~\cite{haldane1991}. Moreover, 
$S=1/2$ objects are elementary excitations also in gapped systems, as in the case of the Majumdar-Ghosh point of the 
frustrated $J_1-J_2$ Heisenberg model (where the ratio between the first-neighbor coupling $J_1$ and the second-neighbor 
one $J_2$ is equal to $2$). Here, the ground state is doubly degenerate (with long-range dimer order)~\cite{majumdar1969} 
and elementary excitations can be seen as propagating defect boundaries between the two ground states (they are analogous 
to solitons)~\cite{shastry1981}. Anyhow, in one-dimensional systems, fractional excitations are ubiquitous and represent
the general feature of spin models. By contrast, in two spatial dimensions, neat examples of spin liquids are rarer,
and the possibility to have free (i.e., deconfined) spinons when magnetic order is destroyed is not taken for granted.
A beautiful example in which fractional excitations are present is given by the Kitaev (compass) model on the honeycomb
lattice~\cite{kitaev2006}, where elementary excitations are Majorana fermions and $Z_2$ gauge fluxes.

In the last 20 years, a huge effort has been devoted to assessing the low-energy behavior of various frustrated spin models,
in order to unveil possible spin-liquid ground states. Many different lattice structures and various kinds of interactions,
including long-range superexchange and ring-exchange terms, have been considered by employing a large variety of analytical
and numerical techniques~\cite{lacroixbook}. Nevertheless, most of these studies focused on the ground-state properties
by computing correlation functions or different quantities that may give information about the presence or absence of a spin
gap. The direct computation of the spin gap has been performed in a few cases, notably for the Heisenberg model on the 
kagome lattice where this issue has been addressed using the density-matrix renormalization group~\cite{yan2011,depenbrock2012}, 
variational Monte Carlo~\cite{iqbal2014}, and Lanczos~\cite{lauchli2016} approaches. By contrast, only a few works, mainly 
based upon analytical or semi-analytical approaches, have focused on all the dynamical properties of frustrated spin 
models~\cite{fuhrman2012,mourigal2013,zhitomirsky2013,ghioldi2018}. In this respect, an important quantity that gives 
direct access to the nature of the excitation spectrum is the dynamical structure factor:
\begin{equation}\label{eq:dsf}
S^{a}(q,\omega) = \sum_{\alpha} |\langle \Upsilon_{\alpha}^q | S^{a}_q | \Upsilon_0 \rangle|^2 \delta(\omega-E_{\alpha}^q+E_0),
\end{equation}
where $|\Upsilon_0\rangle$ is the ground state of the system with energy $E_0$, $\{|\Upsilon_{\alpha}^q \rangle\}$ are the excited 
states with momentum $q$ (relative to the ground state) and energy $E_{\alpha}^q$, and 
\begin{equation}
S^{a}_q=\frac{1}{\sqrt{L}} \sum_R e^{iqR} S^{a}_R
\end{equation}
is the Fourier transformed spin operator for the components $a=(x,y,z)$, with $L$ being the number of sites of the cluster. 
In this regard, inelastic neutron scattering provides a direct measurement of the dynamical structure factor as a function 
of momentum $q$ and energy $\omega$. Therefore, the theoretical computation of $S^{a}(q,\omega)$ has an immediate connection 
to experiments, thus validating or disproving the modelization of a given real material~\cite{coldea2001,ronnow2001,lake2013}.

Unfortunately, the theoretical evaluation of the dynamical structure factor is possible only for very limited cases. In one 
dimension, an exact calculation is possible for the Heisenberg model with inverse-square superexchange~\cite{haldane1993},
while very accurate approximations are now available for the simple Heisenberg model with nearest-neighbor 
interaction~\cite{karbach1997,caux2005,caux2006}. In two spatial dimensions, an exact computation of the dynamical
structure factor is possible for the Kitaev model in both the gapless and gapped regimes~\cite{knolle2014}. Besides these 
fortunate cases in which the model is integrable, numerical techniques have been employed to study generic models,
especially in one dimension. Here, exact diagonalizations can be performed on relatively small clusters~\cite{yokoyama1997}
and their results can be compared with semi-analytical calculations~\cite{lavarelo2014}. Moreover, the density-matrix
renormalization group~\cite{barthel2009} or matrix-product states~\cite{vanderstraeten2016,xie2018} can be used. Alternatively, 
a variational technique, implemented within a quantum Monte Carlo method, has been suggested to approximate the exact sprectrum 
with $L$ states for each momentum $q$~\cite{li2010}. The important advantage of this approach is that the dynamical structure 
factor $S(q,\omega)$ is directly accessible, without any computation requiring (unstable) transformations from real or imaginary 
times to frequencies~\cite{notaMPS}. Curiously, this approach has been applied in very few cases~\cite{dallapiazza2015,mei2015}, 
without systematic benchmarks and comparisons with other methods.

In this paper, we employ the variational approach that was introduced in Ref.~\cite{li2010} to compute the dynamical 
structure factor of Eq.~(\ref{eq:dsf}) for momentum $q$ and energy $\omega$ in the spin-$1/2$ $J_1-J_2$ Heisenberg model in
one dimension:
\begin{equation}\label{eq:model}
{\cal H} = J_1 \sum_{R} \mathbf{S}_R \cdot \mathbf{S}_{R+1} + J_2 \sum_{R} \mathbf{S}_R \cdot \mathbf{S}_{R+2},
\end{equation}
where $R=1,2,\dots,L$ are the (integer) coordinates of the $L$ sites and $\mathbf{S}_R=(S^x_R,S^y_R,S^z_R)$ is the $S=1/2$
spin operator on site $R$. Periodic boundary conditions are considered in the spin Hamiltonian. In the following, we will 
consider the case with $a \equiv z$ in Eq.~(\ref{eq:dsf}), since, given the $SU(2)$ symmetry of the Heisenberg model, any component 
of the structure factor gives the same result. The variational Monte Carlo results are compared with Lanczos diagonalizations on a 
small $L=30$ cluster in order to show the accuracy of the method for different values of the frustrating ratio $J_2/J_1$. Then, 
calculations are reported for large systems, illustrating how the various features of the dynamical structure factor evolve from the 
gapless to the gapped phase, also entering in the incommensurate region with $J_2/J_1>0.5$.

The rest of the paper is organized as follows: in sections~\ref{sec:method} and~\ref{sec:wavefunction} we describe the 
variational approach that we have employed; in section~\ref{sec:results}, we present our numerical results; finally, 
in section~\ref{sec:conclusions}, we discuss the conclusions and the perspectives.

%%%%%%%%%%%%%%%%%%%%%%%%%%%%%%%%%%%%%%%%%%%%%
\begin{figure}
\includegraphics[width=1.0\columnwidth]{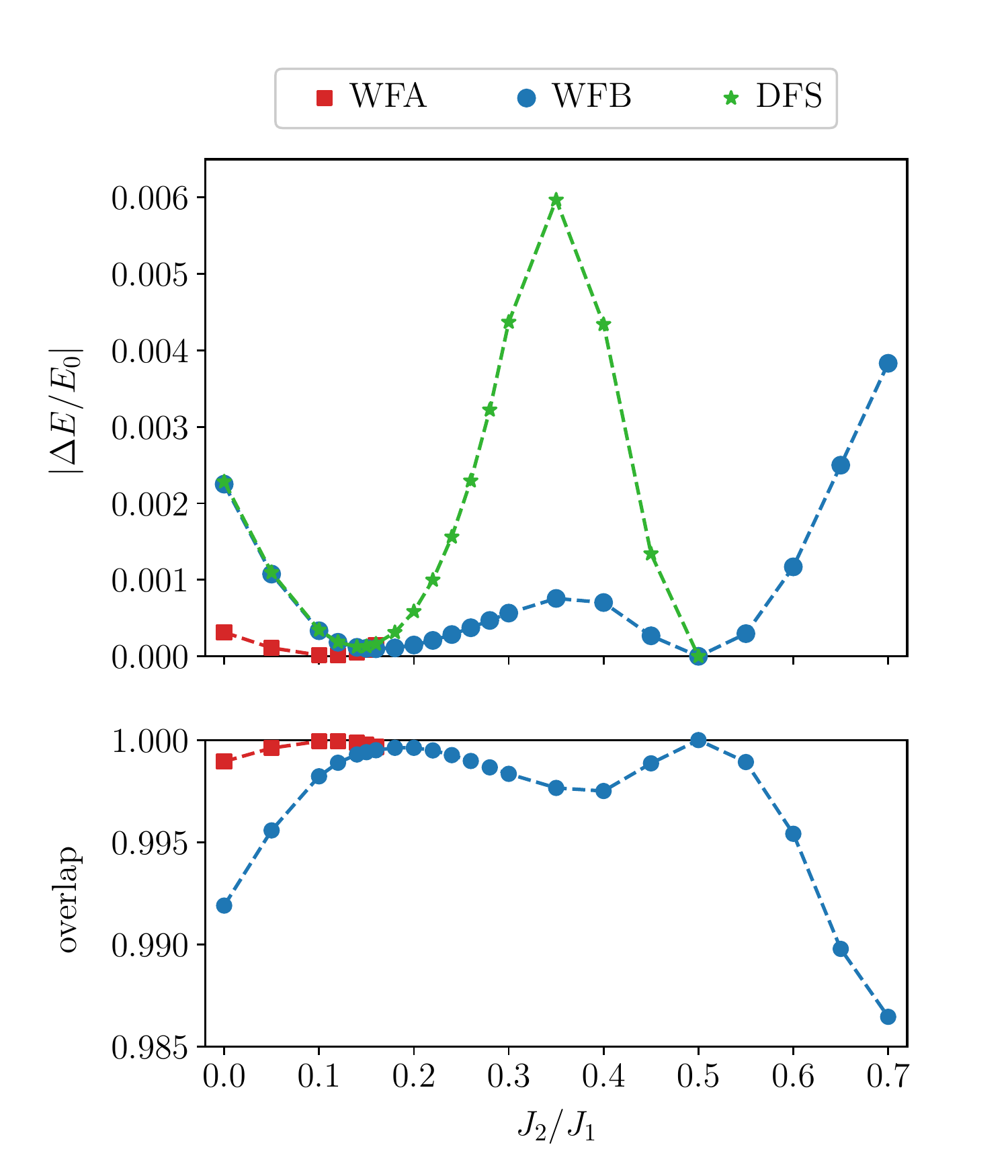}
\caption{\label{fig:accuracyGS}
Upper panel: accuracy of the \textit{DFS}, \textit{WFA}, and \textit{WFB} wave functions for a chain of $L=30$ sites. $\Delta E$ 
is the difference between the variational energy ($E_0^{\rm var}$) and the exact ground-state energy ($E_0$), obtained with Lanczos 
diagonalizations. For $J_2/J_1>0.5$, the accuracy of the \textit{DFS} state is not shown, since it rapidly deteriorates. Lower panel: 
overlap $|\langle \Psi_0 | \Upsilon_0\rangle|$ between the variational wave functions, either \textit{WFA} or \textit{WFB}, and the 
exact one.}
\end{figure}
%%%%%%%%%%%%%%%%%%%%%%%%%%%%%%%%%%%%%%%%%%%%%

\section{Variational method}\label{sec:method}

The variational approach is based on a Gutzwiller projected fermionic wave function, which is constructed from an auxiliary
superconducting (BCS) Hamiltonian:
\begin{eqnarray}
{\cal H}_{0} &=& \sum_{R,R^\prime,\sigma} t_{R,R^\prime} c_{R,\sigma}^\dagger c_{R^\prime,\sigma}^\dagga \nonumber \\
             &+& \sum_{R,R^\prime} \Delta_{R,R^\prime} \left ( c_{R,\uparrow}^\dagger c_{R^\prime,\downarrow}^\dagger + 
             c_{R^\prime,\uparrow}^\dagger c_{R,\downarrow}^\dagger \right ) + H.c.;
\label{eq:H0}
\end{eqnarray}
here, $c_{R,\sigma}^\dagger$ ($c_{R,\sigma}^\dagga$) creates (destroys) an electron with spin $\sigma=\pm 1/2$ on site $R$; 
$t_{R,R^\prime}$ and $\Delta_{R,R^\prime}$ are hopping and singlet pairing terms, respectively. This Hamiltonian is quadratic
in the fermionic operators and, therefore, can be easily diagonalized. Its ground state is denoted by $|\Phi_0\rangle$. 
Of course, this quantum state is not suitable to describe a spin system, since it is defined in the ``enlarged'' Hilbert 
space with also empty and doubly occupied sites. Then, a suitable variational wave function for the Heisenberg model of 
Eq.~(\ref{eq:model}) can be obtained by projecting out all the configurations with at least one empty or doubly occupied
site:
\begin{equation}
|\Psi_0\rangle = \mathcal{P}_G |\Phi_0\rangle,
\end{equation}
where $\mathcal{P}_G=\prod_R(n_{R,\uparrow}-n_{R,\downarrow})^2$ (with $n_{R,\sigma}= c_{R,\sigma}^\dagger c_{R,\sigma}^\dagga$
being the local electron density per spin $\sigma$ on site $R$) is the Gutzwiller projector, which enforces single 
fermionic occupation on each site. It has been shown that Gutzwiller-projected fermionic wave functions are very accurate 
for describing the exact ground state of the one-dimensional Heisenberg model with $J_2=0$, as well as the lowest-energy spinon 
excitations~\cite{yunoki2006}.

%%%%%%%%%%%%%%%%%%%%%%%%%%%%%%%%%%%%%%%%%%%%%
\begin{figure*}
\includegraphics[width=0.65\columnwidth]{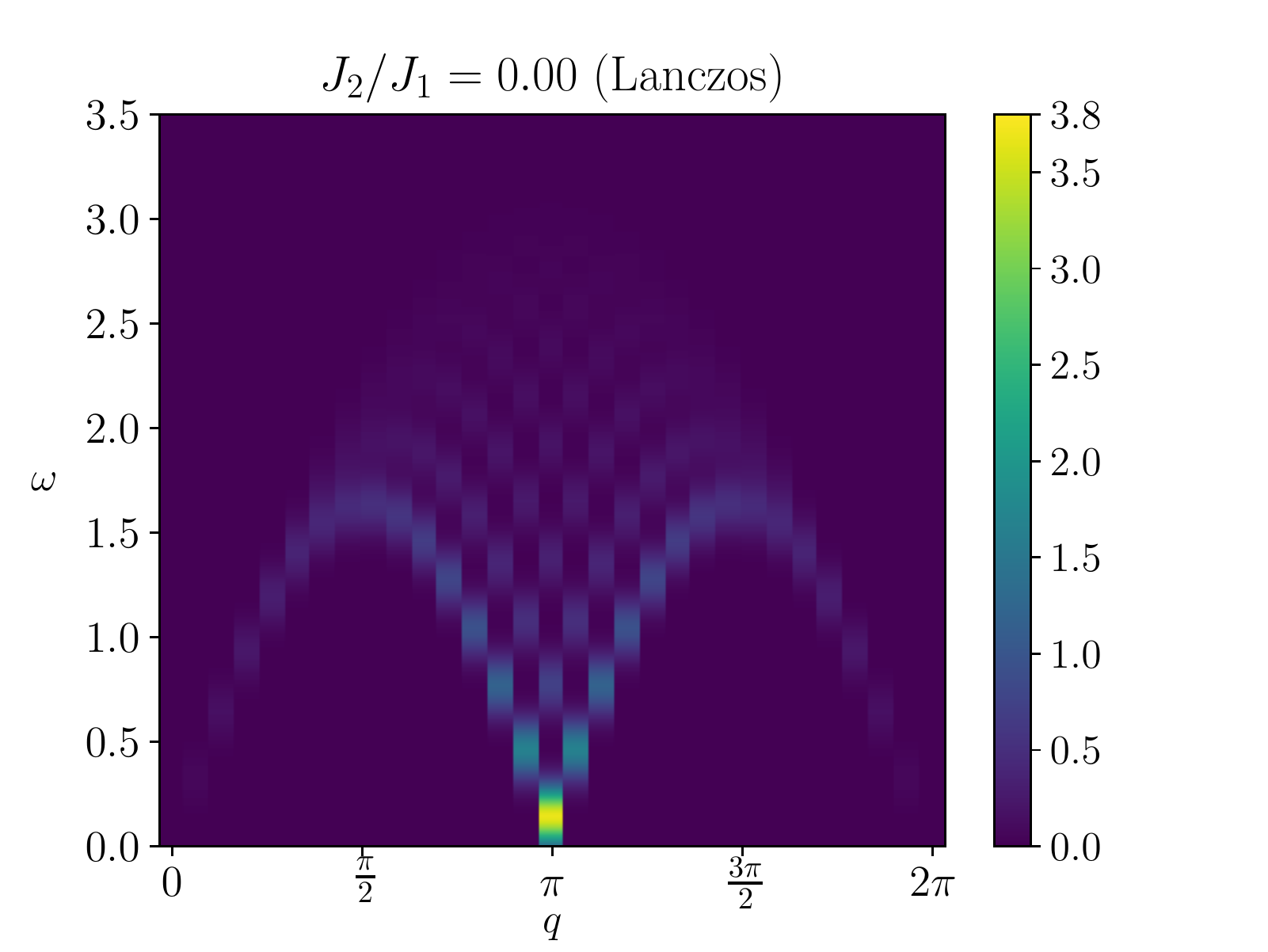}
\includegraphics[width=0.65\columnwidth]{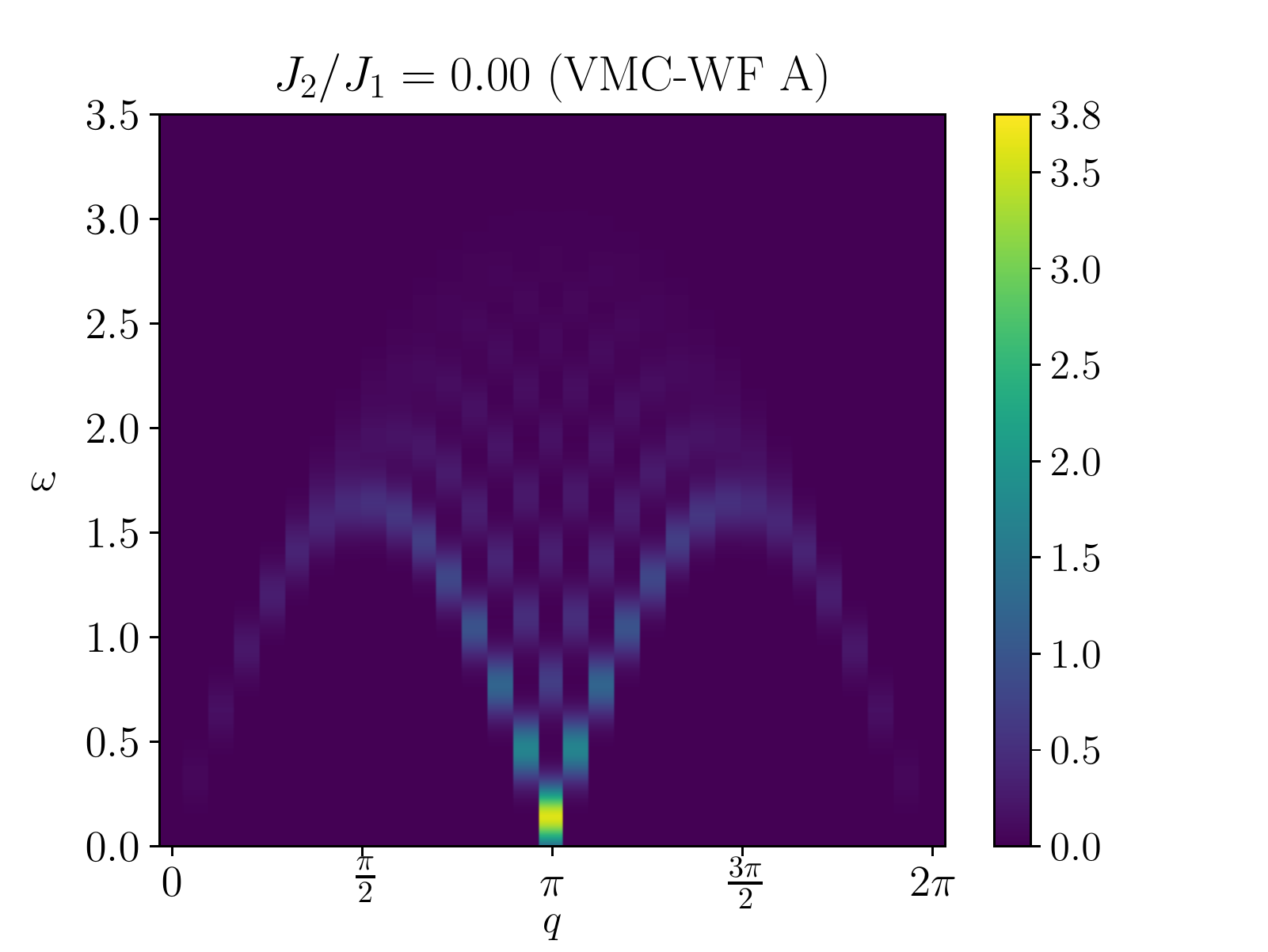}
\includegraphics[width=0.65\columnwidth]{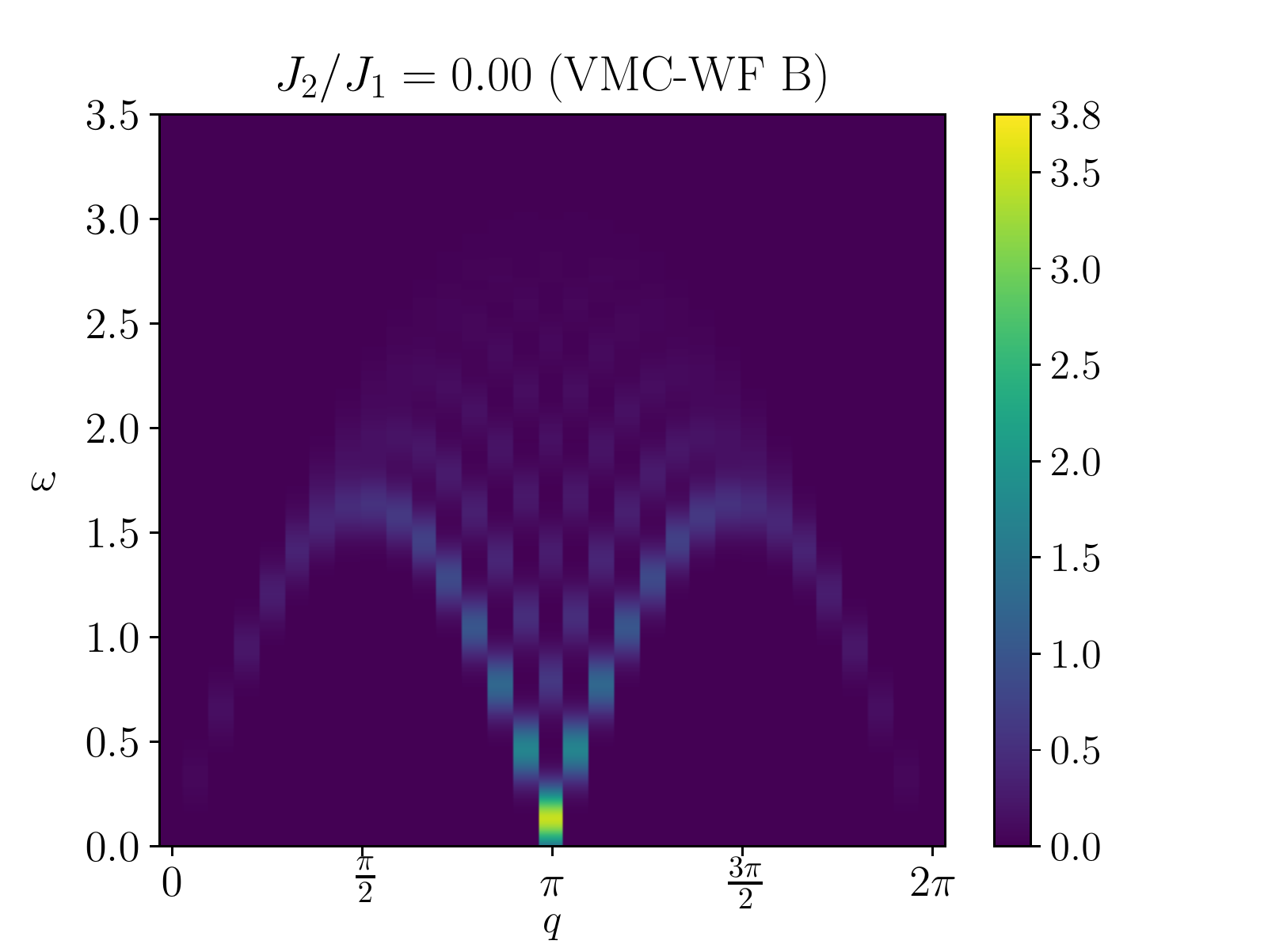}
\caption{\label{fig:vmc_lanczos}
Dynamical structure factor for $L=30$ and $J_2=0$. The Lanczos results are reported in the left panel. The variational calculations 
in the middle and right panels were obtained using the \textit{Ans\"atze} \textit{WFA} and \textit{WFB}, respectively.}
\end{figure*}

\begin{figure}
\includegraphics[width=1.0\columnwidth]{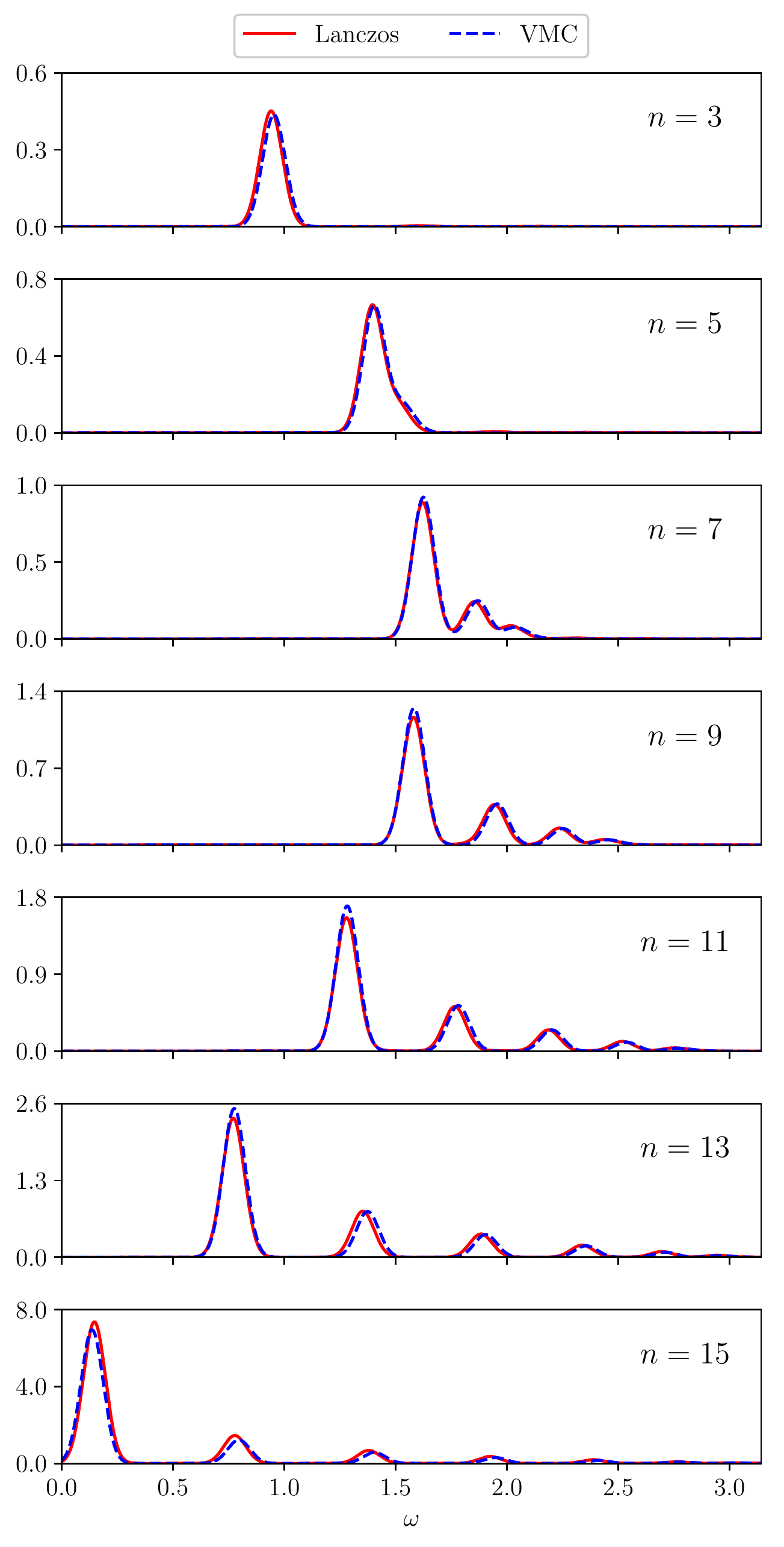}
\caption{\label{fig:lanczos1}
Comparison between Lanczos and variational calculations for $S^{z}(q,\omega)$ at different momenta $q=2\pi/L \times n$, with $n$ being an 
integer specified in the figure. Here, we consider $L=30$ and $J_2=0$. The delta-functions in Eqs.~(\ref{eq:dsf}) and~(\ref{eq:Szz_practical}) 
have been replaced by normalized Gaussians with $\sigma=0.05 J_1$. Statistical errors are negligible within the present scale.}
\end{figure}

\begin{figure}
\includegraphics[width=1.0\columnwidth]{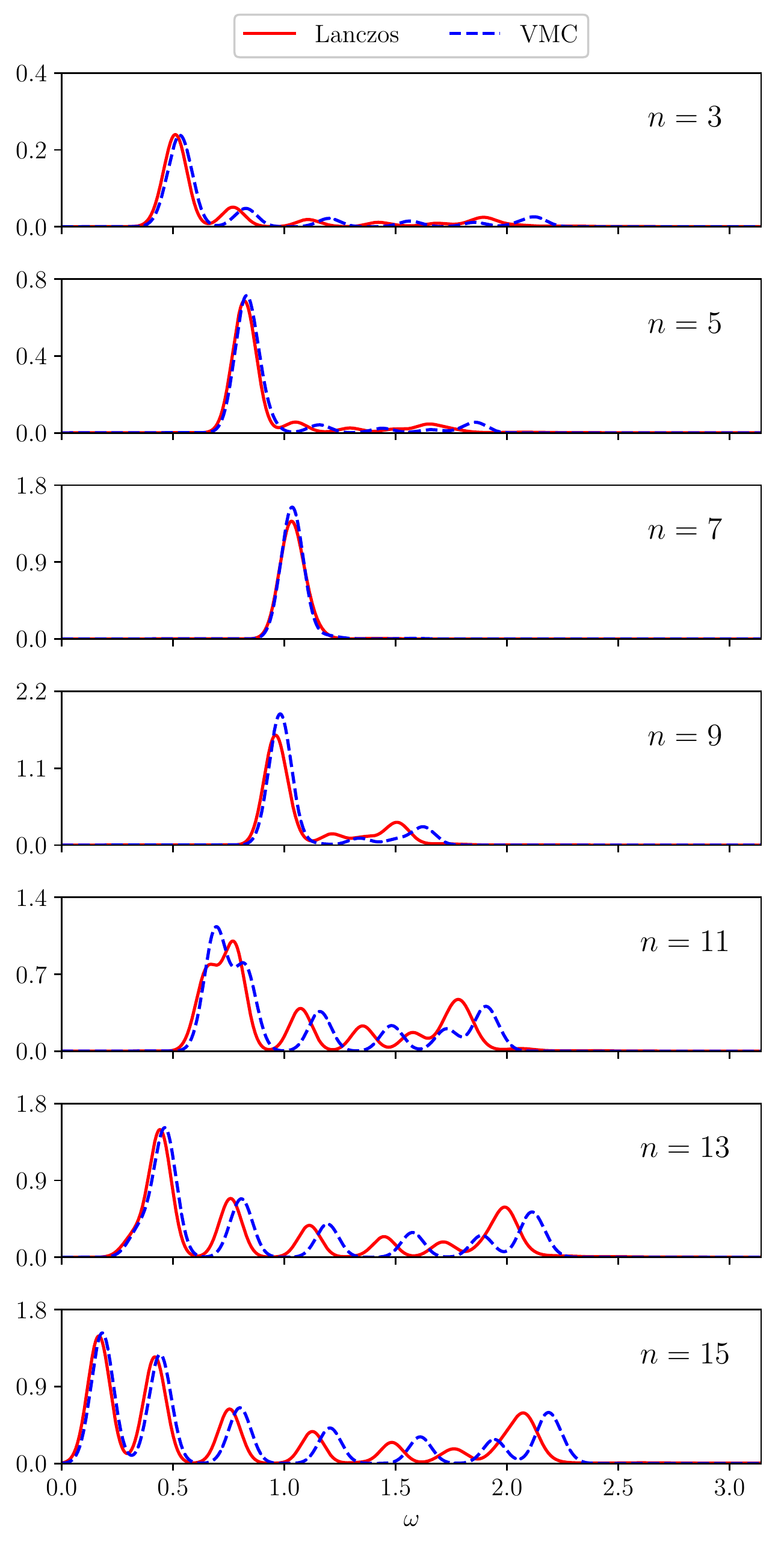}
\caption{\label{fig:lanczos2}
The same as in Fig.~\ref{fig:lanczos1} but for $J_2/J_1=0.45$.}
\end{figure}
%%%%%%%%%%%%%%%%%%%%%%%%%%%%%%%%%%%%%%%%%%%%%

The parameters of ${\cal H}_{0}$, i.e., hopping and pairing amplitudes, are taken to be real and fully optimized by means 
of the stochastic reconfiguration technique, in order to minimize the variational energy of $|\Psi_0\rangle$~\cite{sorella2005}.
In most of the calculations, we will impose translational symmetry in the quadratic Hamiltonian ${\cal H}_{0}$, strongly
reducing the number of independent parameters to be treated. We must emphasize the fact that both periodic boundary conditions 
(PBC) and anti-periodic boundary conditions (APBC) are allowed within the auxiliary BCS Hamiltonian (leading to a real wave 
function). However, while in the presence of a gapped fermionic spectrum either options will lead to a \textit{unique} ground state, 
the same may not be true for a gapless spectrum. For example, if there are gapless points at $k=\pm \pi/2$, the ground state is 
unique if PBC (APBC) are considered for $L=4n+2$ ($L=4n$), where $n$ is an integer.

In order to tackle the problem of computing $S^{z}(q,\omega)$, we need to devise a way to construct excited states. Following 
the procedure of Ref.~\cite{li2010}, we first introduce a set of {\it two-spinon} triplet excitations with momentum $q$, which 
are obtained by the Gutzwiller projection of particle-hole fermionic excitations:
\begin{equation}\label{eq:qRstate}
|q,R\rangle =  \mathcal{P}_G \frac{1}{\sqrt{L}} \sum_{R^\prime} e^{iqR^\prime} \sum_{\sigma} \sigma 
c^\dagger_{R+R^\prime,\sigma}c^\dagga_{R^\prime,\sigma} |\Phi_0\rangle.
\end{equation}
Then, for each momentum $q$, $\{|q,R\rangle \}$ defines a (non-orthogonal) basis set that can be used to approximate the exact
low-energy excitations. In other words, we can define a set of $L$ states~\cite{note1} (for each momentum $q$), which are labeled by $n$:
\begin{equation}\label{eq:var_excitations}
|\Psi_n^q\rangle = \sum_R A^{n,q}_R |q,R\rangle.
\end{equation}
The coefficients $A^{n,q}_R$ are obtained by diagonalizing the Heisenberg Hamiltonian within the subspace generated by 
$\{|q,R\rangle \}$ for each $q$, namely by solving the generalized eigenvalue problem:
\begin{equation}\label{eq:general_eig_prob}
\sum_{R^\prime} H^q_{R,R^\prime}  A^{n,q}_{R^\prime} = E_n^q \sum_{R^\prime} O^q_{R,R^\prime} A^{n,q}_{R^\prime},
\end{equation}
where we have introduced two matrices, $H^q_{R^\prime,R}=\langle q,R^\prime|{\cal H}|q,R \rangle$ (Hamiltonian) and 
$O^q_{R^\prime,R}=\langle q,R^\prime|q,R \rangle$ (overlap). Notice that, within our fermionic variational approach, the basis set
$\{|q,R\rangle \}$ provides a natural generalization of the well-known {\it single-mode approximation}~\cite{auerbachbook} that 
is recovered by restricting ourselves to consider only $|q,0\rangle= S^z_q |\Psi_0\rangle$. 

Finally, the dynamical structure factor of Eq.~(\ref{eq:dsf}) 
is approximated by taking:
\begin{equation}\label{eq:Szz_practical}
S^{z}(q,\omega) = \sum_n |\langle \Psi_{n}^q | S^{z}_q | \Psi_0 \rangle|^2 \delta(\omega-E_{n}^q+E_0^{\rm var}),
\end{equation}
where, compared to the exact form of Eq.~(\ref{eq:dsf}), the variational states $| \Psi_0 \rangle$ and 
$\{| \Psi_{n}^q \rangle \}$ are considered (instead of the exact eigenstates), and the variational energies 
$E_0^{\rm var}$ (corresponding to $|\Psi_0 \rangle$) and $\{ E_{n}^q \}$ are taken (instead
of the exact ones). Most importantly, the sum over excited states runs over at most $L$ states (instead of an exponentially 
large number). By using Eq.~(\ref{eq:var_excitations}), we have:
\begin{equation}\label{eq:Szz_practical2}
S^{z}(q,\omega) = \sum_n \left|\sum_R (A^{n,q}_R)^* O^q_{R,0}\right|^2 \delta(\omega-E_n^q+E_0^{\rm var}).
\end{equation}

Remarkably, all the quantities that define the dynamical structure factor of Eq.~(\ref{eq:Szz_practical2}) can be computed 
within a variational Monte Carlo scheme (i.e., without any sign problem). In fact, the entries of the Hamiltonian and overlap
matrices are given by:
\begin{eqnarray}
H^q_{R^\prime,R} &=& \sum_x \langle q,R^\prime|x \rangle \langle x| {\cal{H}} |q,R\rangle, \\
O^q_{R^\prime,R} &=& \sum_x \langle q,R^\prime|x \rangle \langle x|q,R\rangle,
\end{eqnarray}
where $\{ |x\rangle \}$ is a set of normalized and orthogonal states, which can be sampled by using the variational wave 
function $|\langle x|\Psi_0\rangle|^2$ as the probability distribution:
\begin{eqnarray}
H^q_{R^\prime,R} &=& \sum_x \left[\frac{\langle q,R^\prime|x \rangle}{\langle \Psi_0|x \rangle}
\frac{\langle x| {\cal{H}} |q,R\rangle}{\langle x| \Psi_0 \rangle}\right] |\langle x|\Psi_0 \rangle|^2, \\
O^q_{R^\prime,R} &=& \sum_x \left[\frac{\langle q,R^\prime|x \rangle}{\langle \Psi_0|x \rangle}
\frac{\langle x |q,R\rangle}{\langle x| \Psi_0 \rangle}\right] |\langle x|\Psi_0 \rangle|^2.
\end{eqnarray}
At this stage, it is worth making two remarks. First of all, our sampling procedure is possible because both the ground-state 
wave function $|\Psi_0 \rangle$ and the particle-hole excitations of Eq.~(\ref{eq:qRstate}) have $S^{z}_{\rm tot}=\sum_{R} S^{z}_{R}=0$
and, therefore, the set of configurations $\{ |x\rangle \}$ can be chosen to also have $S^{z}_{\rm tot}=0$. This would not be
possible whenever considering excitations involving a spin flip. For this case, a different sampling procedure was proposed in 
Ref.~\cite{li2010}. The most important advantage of our approach is that all the values of the momentum $q$ can be computed with 
a {\it single} Monte Carlo simulation, at variance with the previous technique, in which each $q$ needs a separate calculation.
Second, within our formulation, the sampling is correct whenever the ground-state wave function $\langle x|\Psi_0\rangle$ is 
nonzero for all the configurations $|x\rangle$; otherwise, the sampling procedure neglects the contributions from these 
``vanishing'' configurations. We checked that, for most of the cases we considered, the number of these configurations is 
negligible and, therefore, they do not affect the final results.

We emphasize that, within this procedure, once the ground-state wave function is optimized, the only remaining parameters are the 
coefficients $\{ A^{n,q}_R \}$, which are completely determined by solving Eq.~(\ref{eq:general_eig_prob}). In other words, the 
particle-hole excitations are applied to a {\it fixed} reference state, i.e., $|\Phi_0\rangle$, which is optimized, once for all,
to minimize the ground-state variational energy.

The evaluation of $O^q_{R^\prime,R}$ and $H^q_{R^\prime,R}$ essentially boils down to computing the following quantities:
\begin{eqnarray}
G_{R,R^\prime}^\sigma(x) &=& \frac{\langle x| c^\dagger_{R,\sigma}c^\dagga_{R^\prime,\sigma}|\Psi_0 \rangle}{\langle x|\Psi_0 \rangle}, \\
 \Gamma_{R,R'}^\sigma(x) &=& \frac{\langle x| {\cal{H}} c^\dagger_{R,\sigma}c^\dagga_{R^\prime,\sigma} |\Psi_0 \rangle}{\langle x|\Psi_0 \rangle}.
\end{eqnarray}
Once $O^q_{R^\prime,R}$ and $H^q_{R^\prime,R}$ have been evaluated, the generalized eigenvalue problem~(\ref{eq:general_eig_prob})
for the excited states of momentum $q$ can be solved. In doing so, we need to get rid of the linear dependence, which may affect the 
set $\{|q,R\rangle\}$. This is achieved by restricting Eq.~(\ref{eq:general_eig_prob}) to the subspace of eigenvectors of the overlap
matrix that have non-zero eigenvalues. In practice, since the entries $O^q_{R^\prime,R}$ are computed stochastically, we need to 
discard the eigenvectors whose eigenvalues are smaller than some given tolerance.

%%%%%%%%%%%%%%%%%%%%%%%%%%%%%%%%%%%%%%%%%%%%%
\begin{figure*}
\includegraphics[width=1.0\columnwidth]{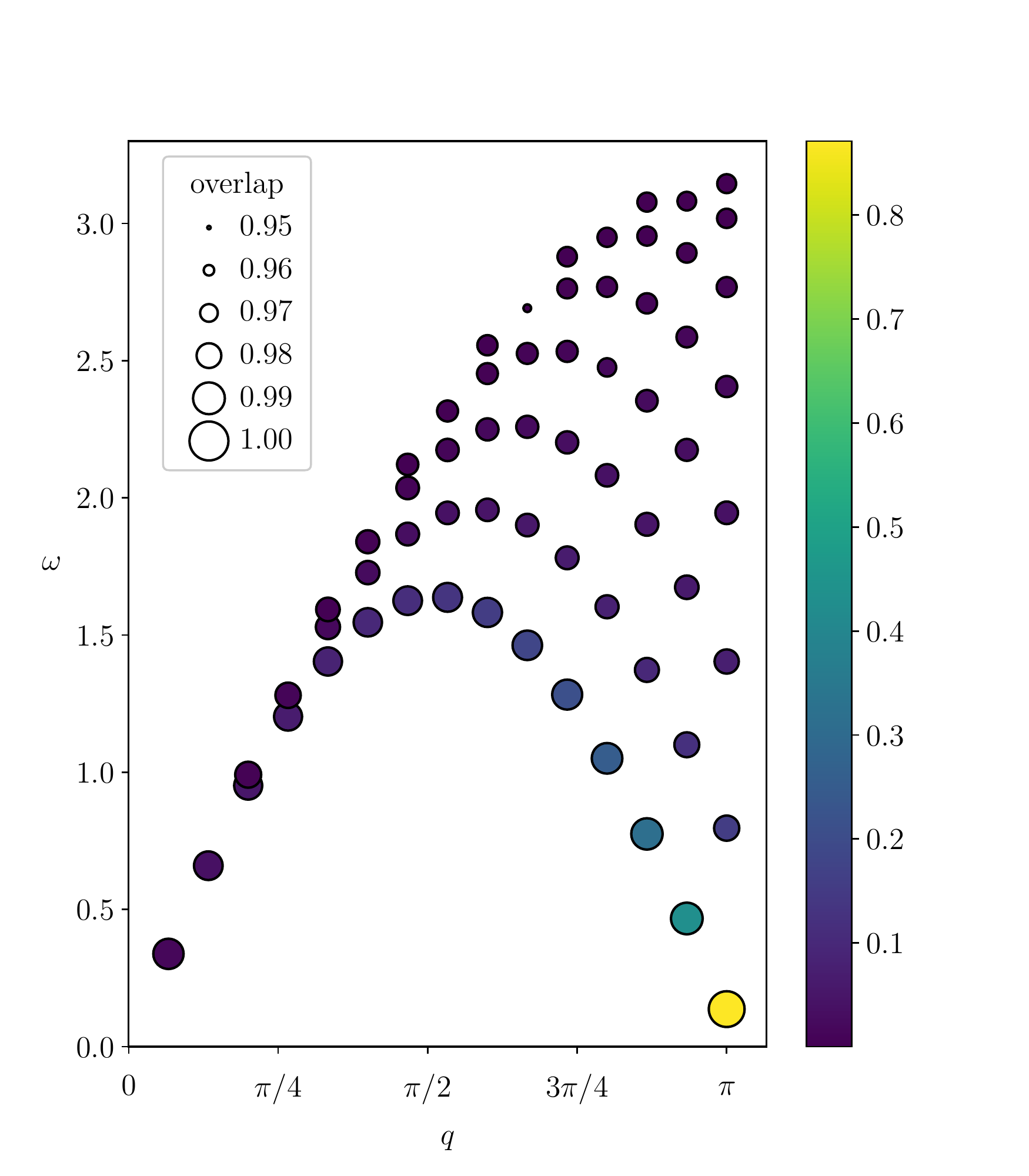}
\includegraphics[width=1.0\columnwidth]{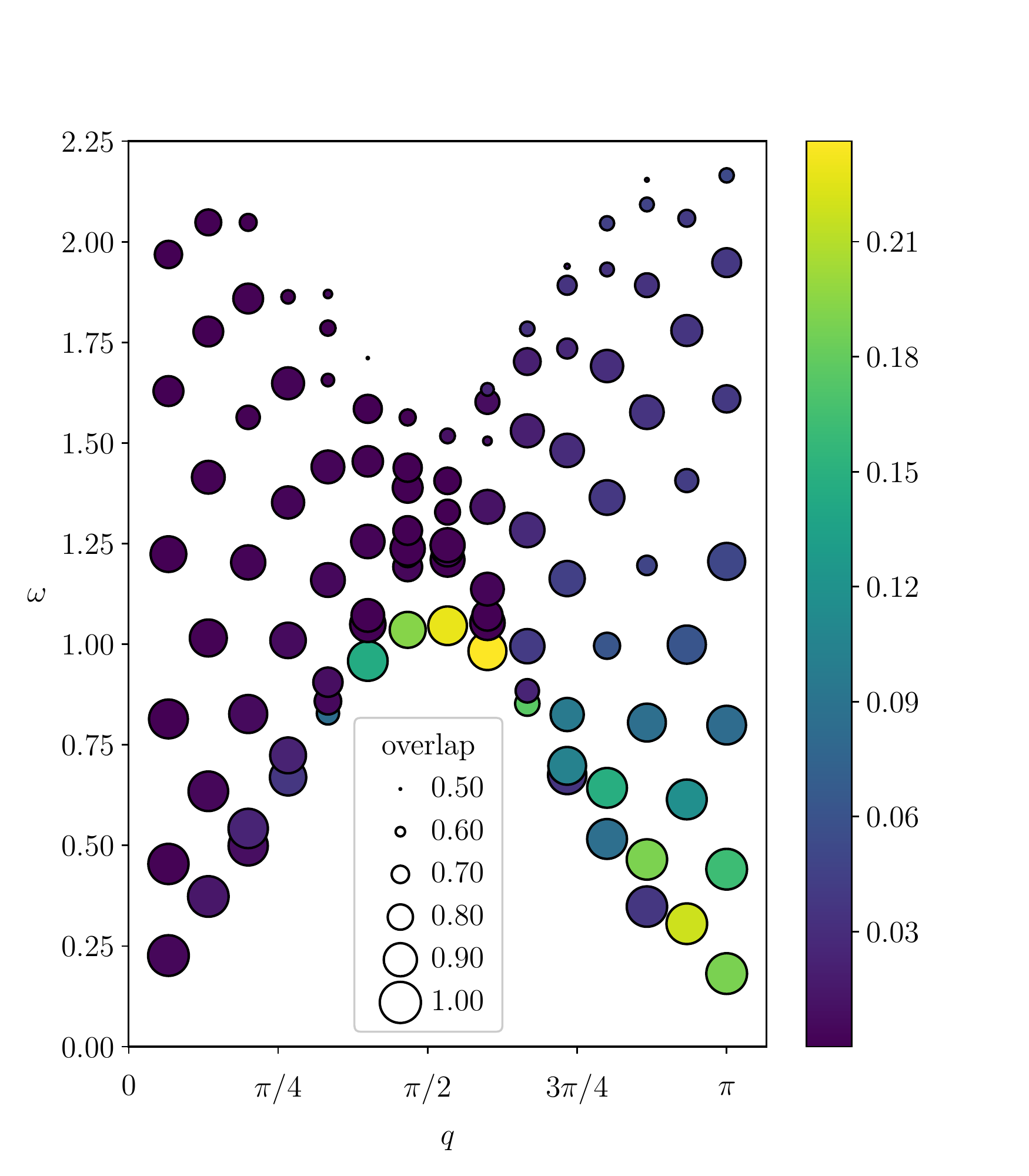}
\caption{\label{fig:accuracy_exc}
The overlaps between the exact and the variational excitations are reported for a chain of $L=30$ sites. The cases with $J_2=0$ (left panel)
and $J_2/J_1=0.45$ (right panel) are shown. The size of the circles indicates the magnitude of the overlap, while their colors represent the
value of the variational spectral weights $|\langle \Psi_{n}^q|S^{z}_q|\Psi_0 \rangle|^2$. For a detailed description of the calculations 
of the overlaps, see the main text.}
\end{figure*}

\begin{figure*}
\includegraphics[width=0.4\textwidth]{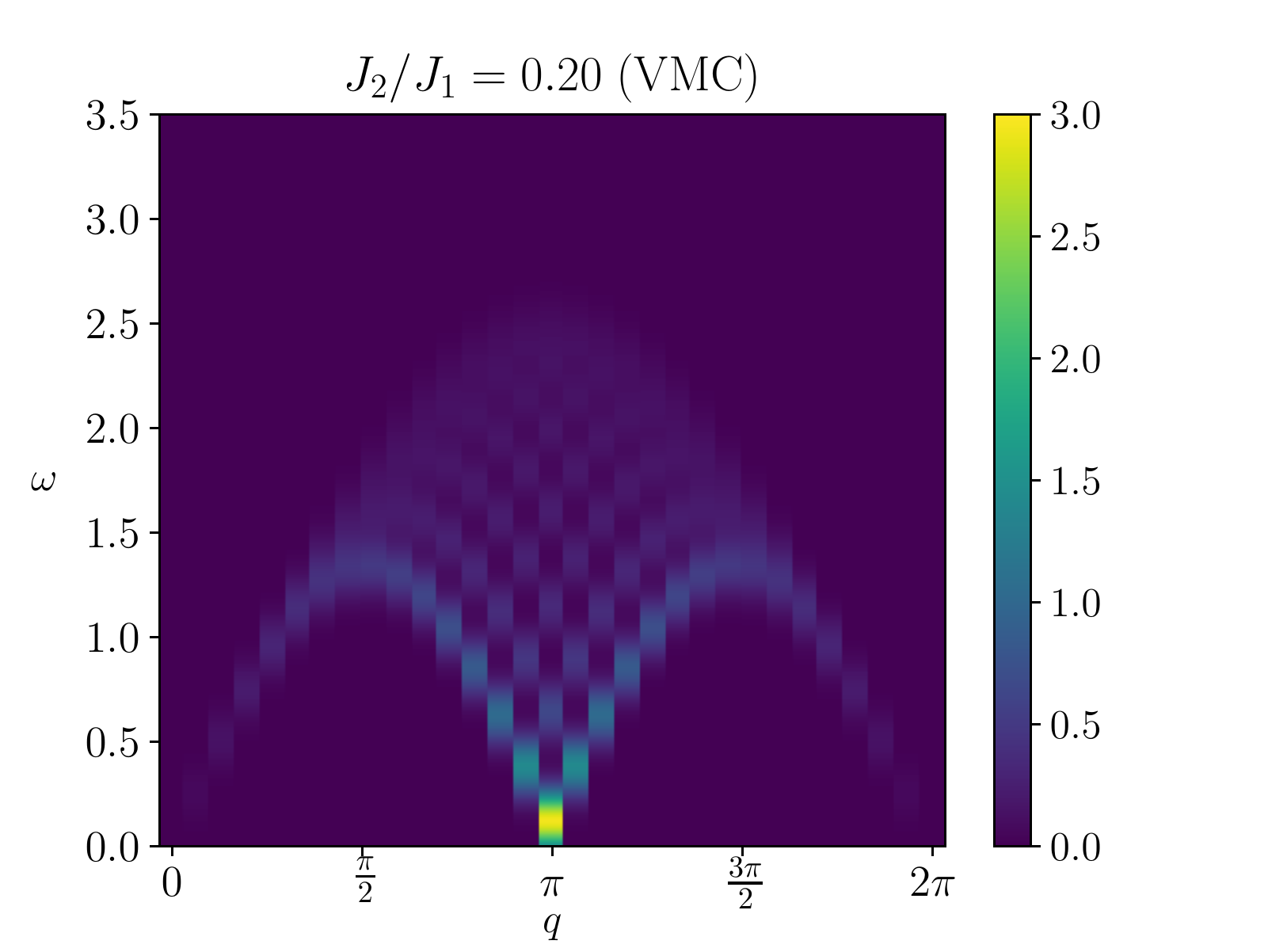}
\includegraphics[width=0.4\textwidth]{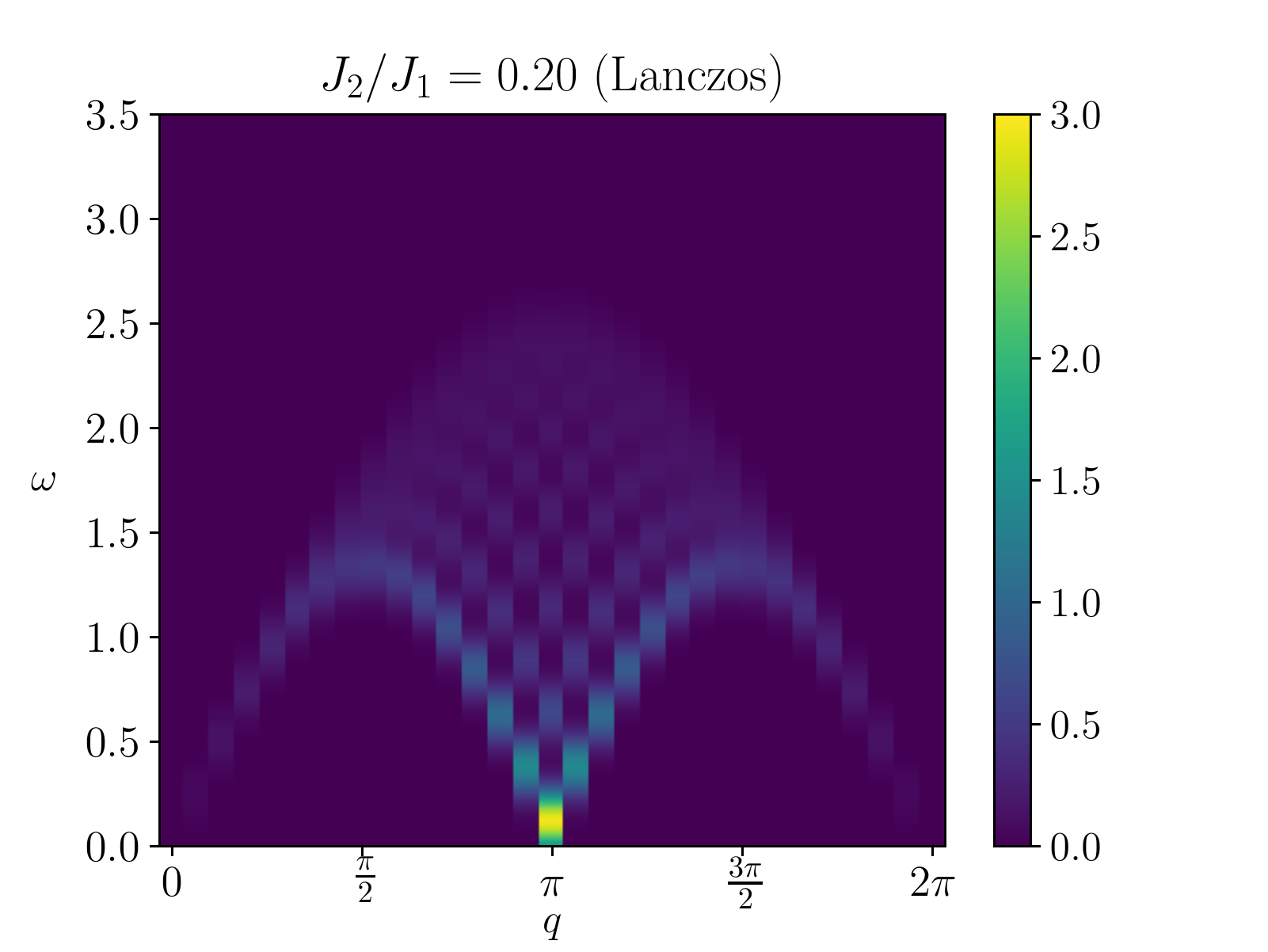}
\includegraphics[width=0.4\textwidth]{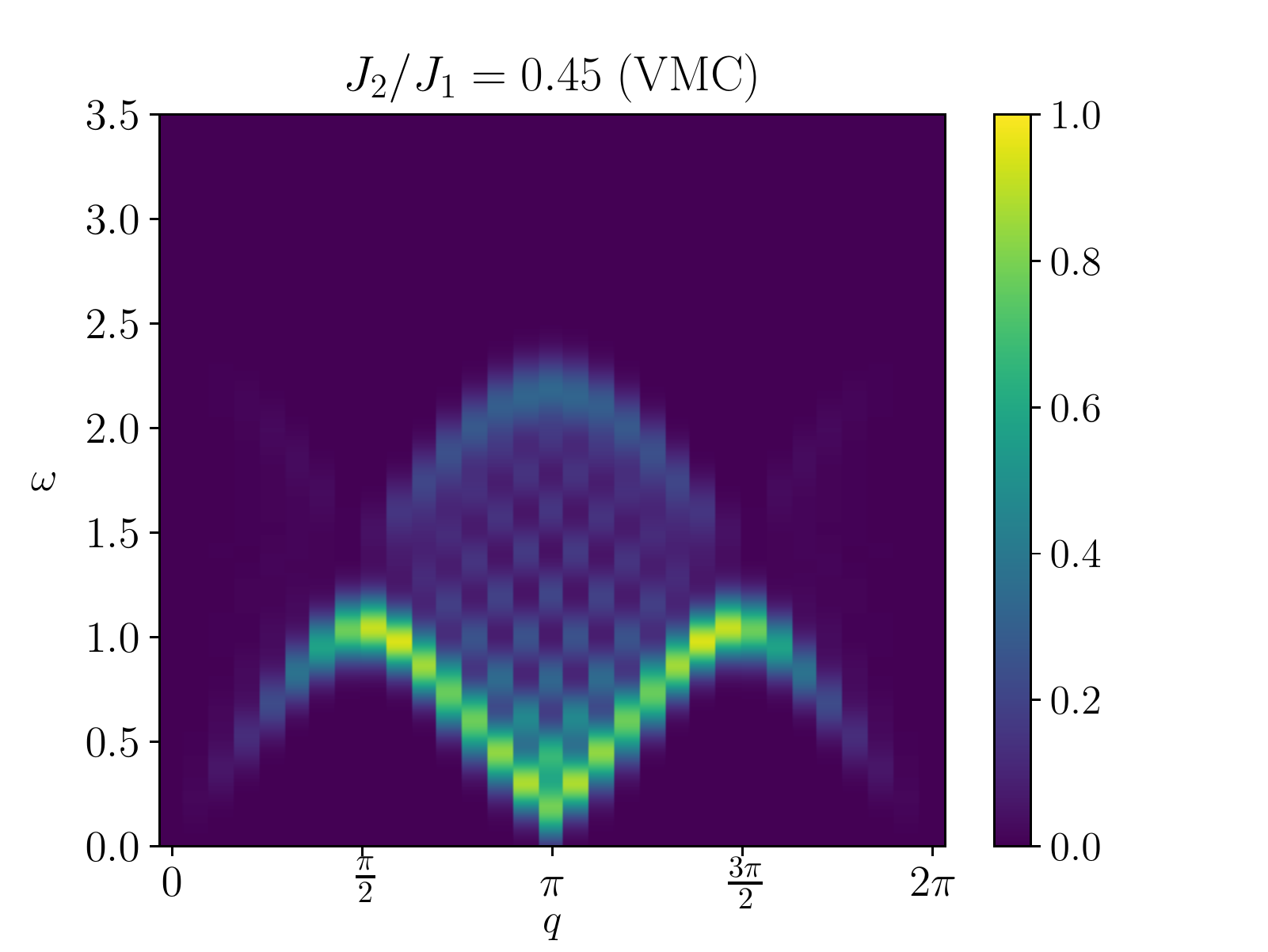}
\includegraphics[width=0.4\textwidth]{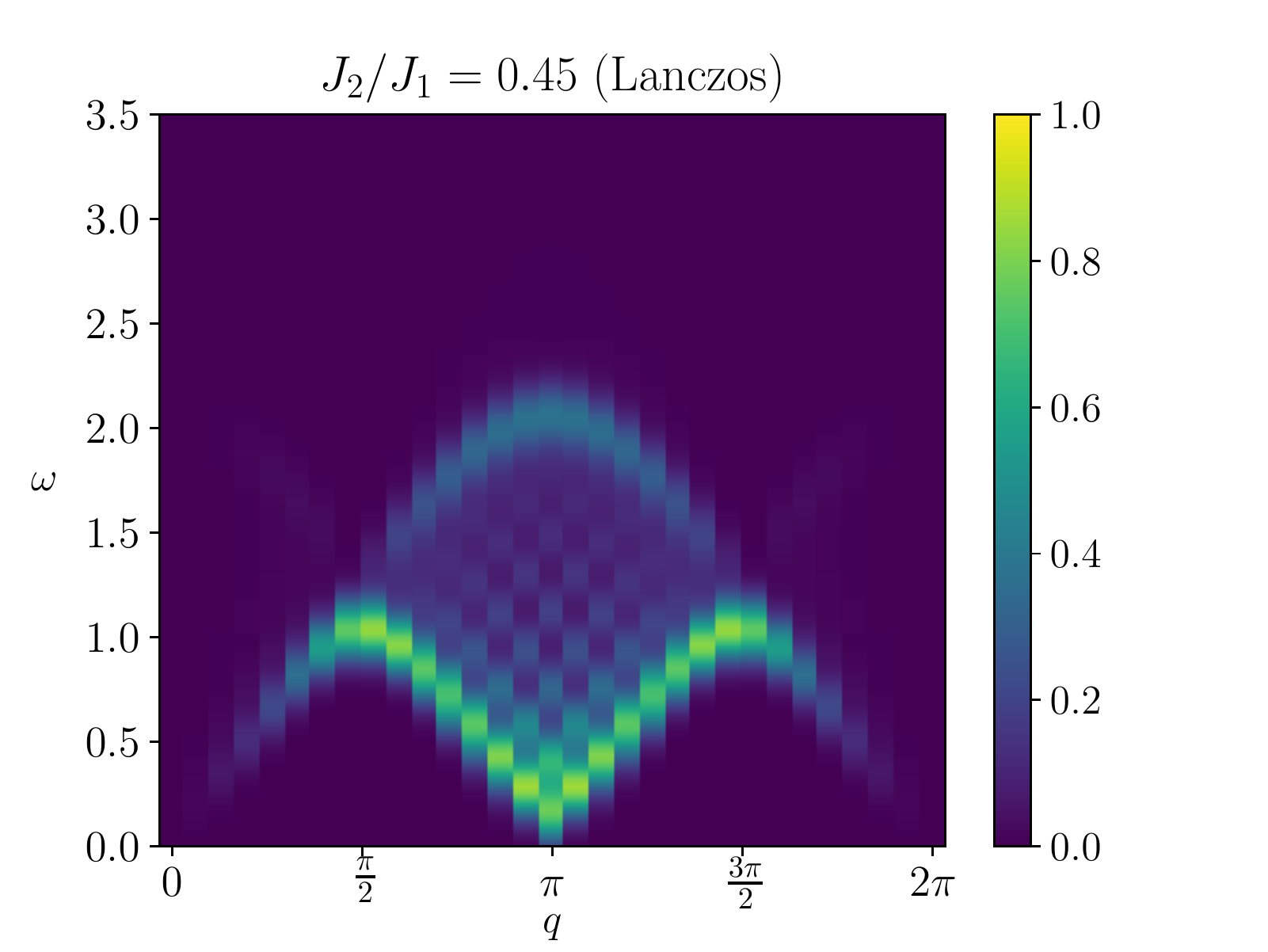}
\includegraphics[width=0.4\textwidth]{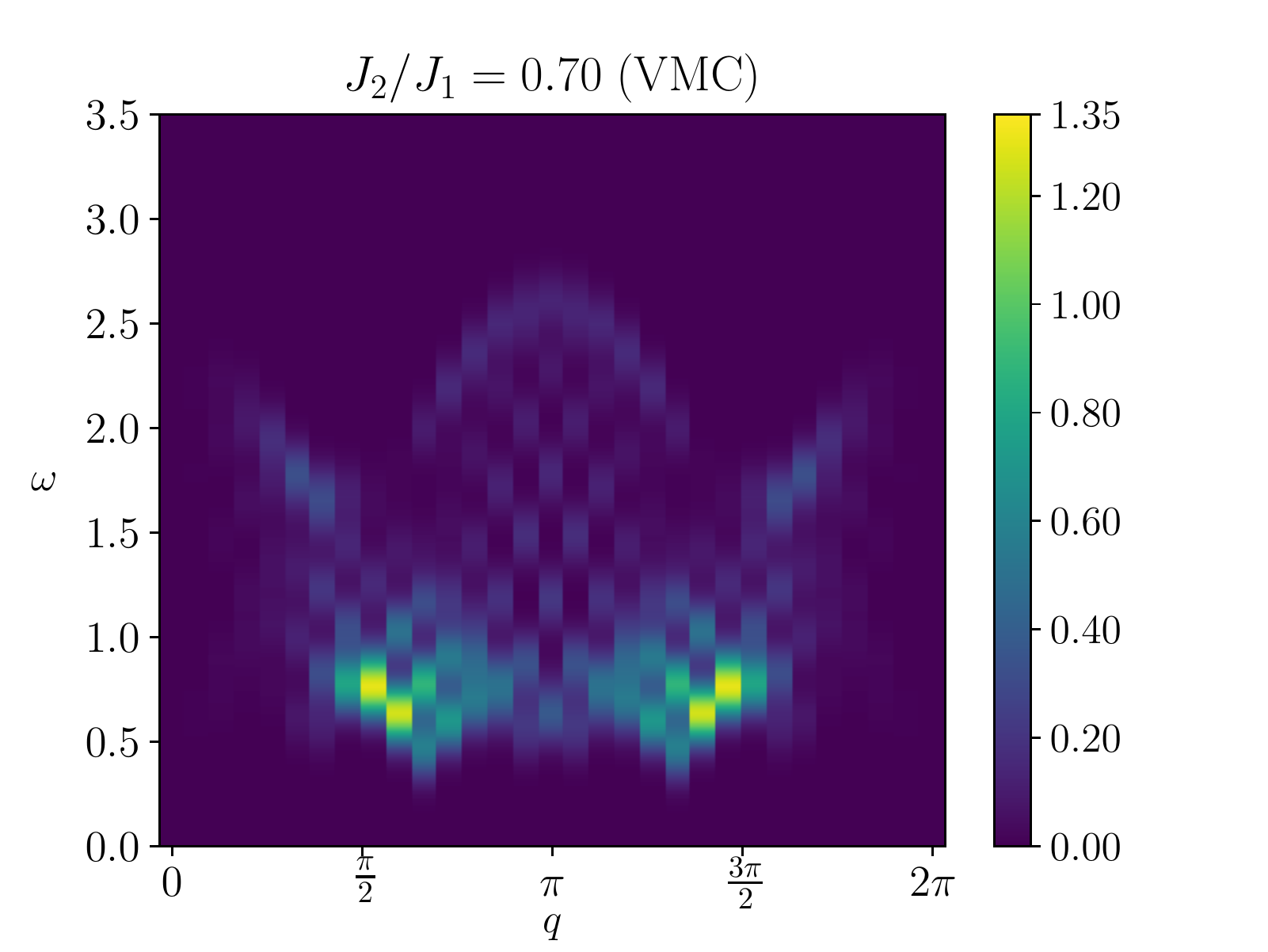}
\includegraphics[width=0.4\textwidth]{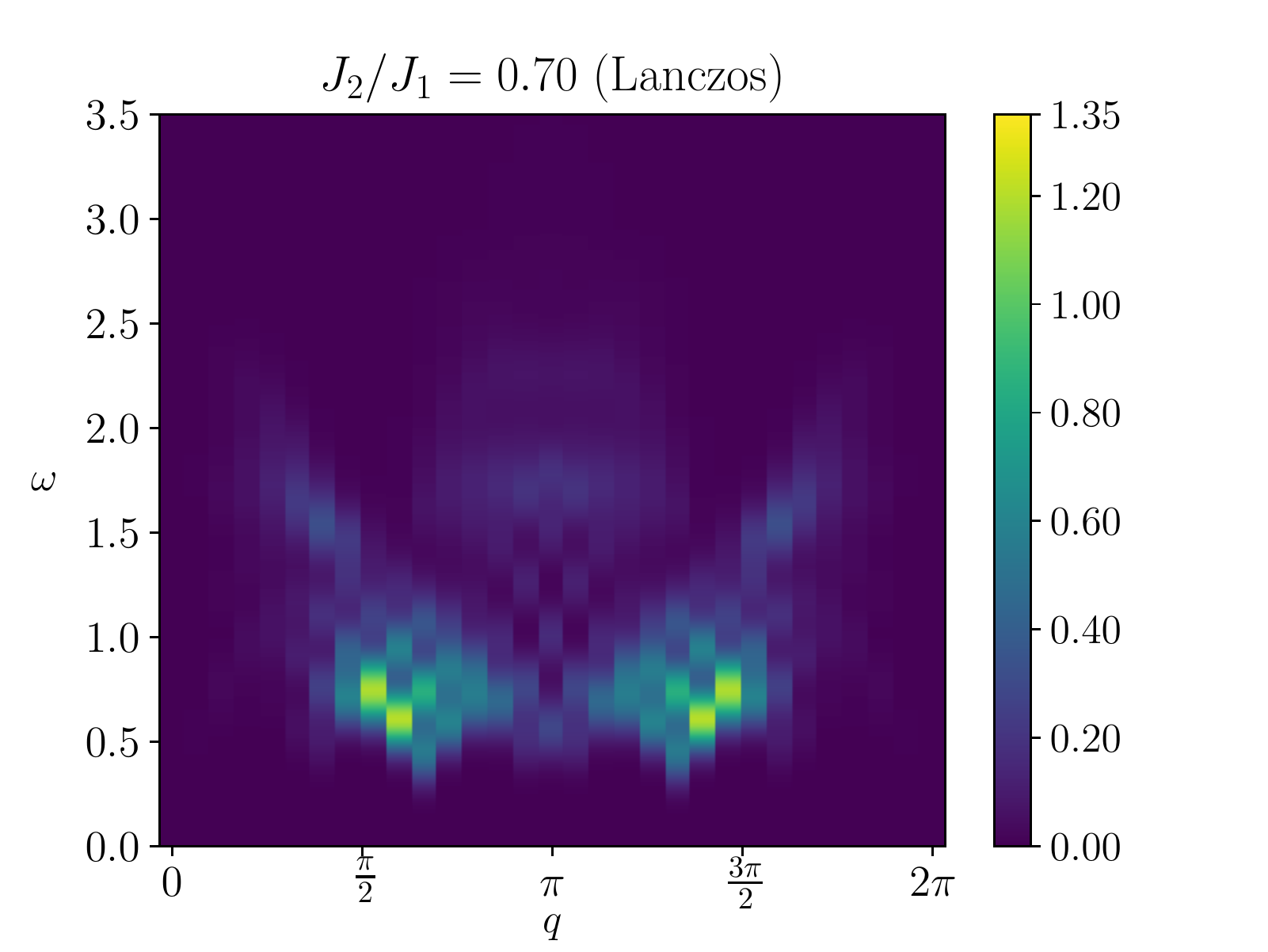}
\includegraphics[width=0.4\textwidth]{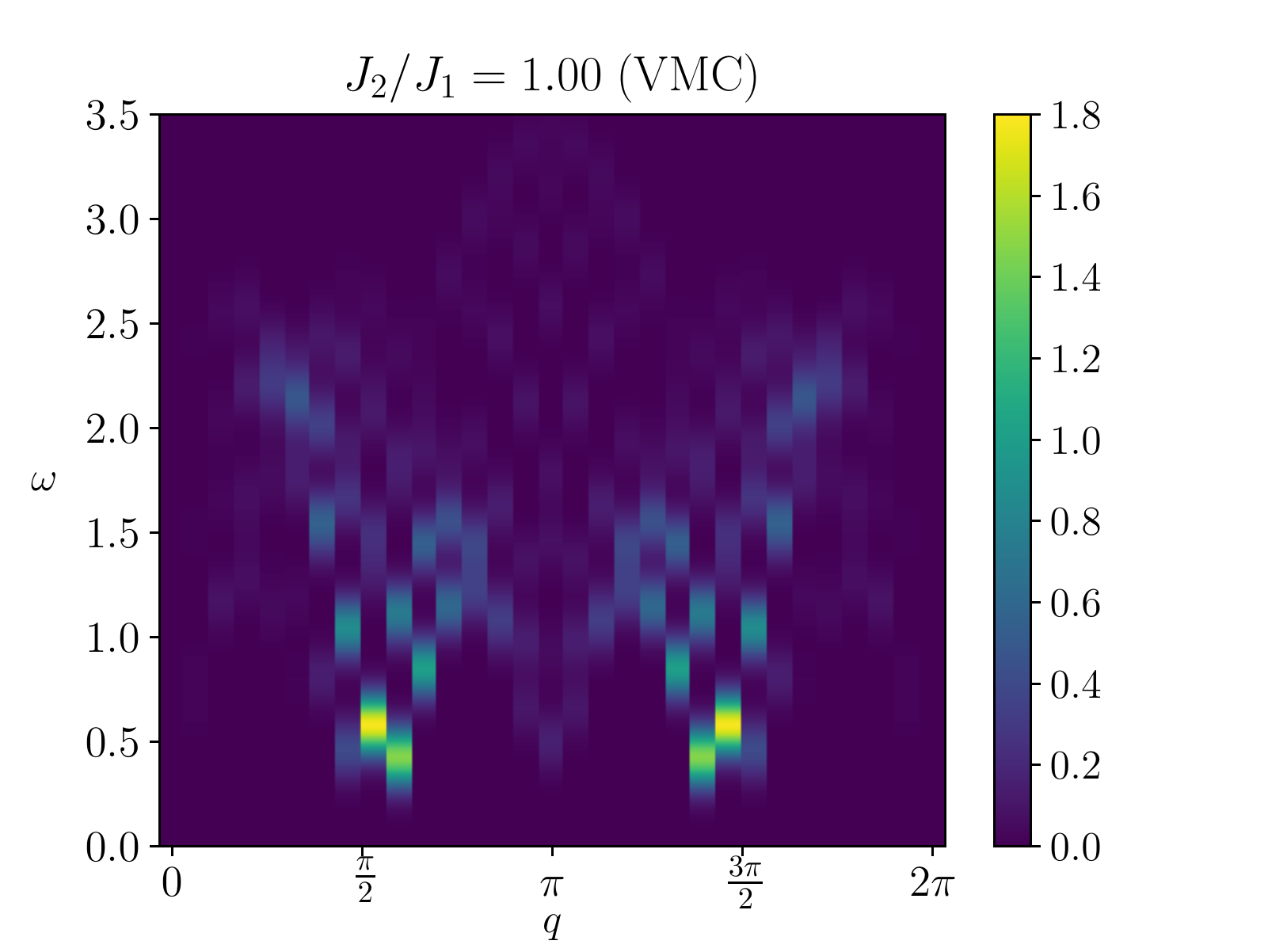}
\includegraphics[width=0.4\textwidth]{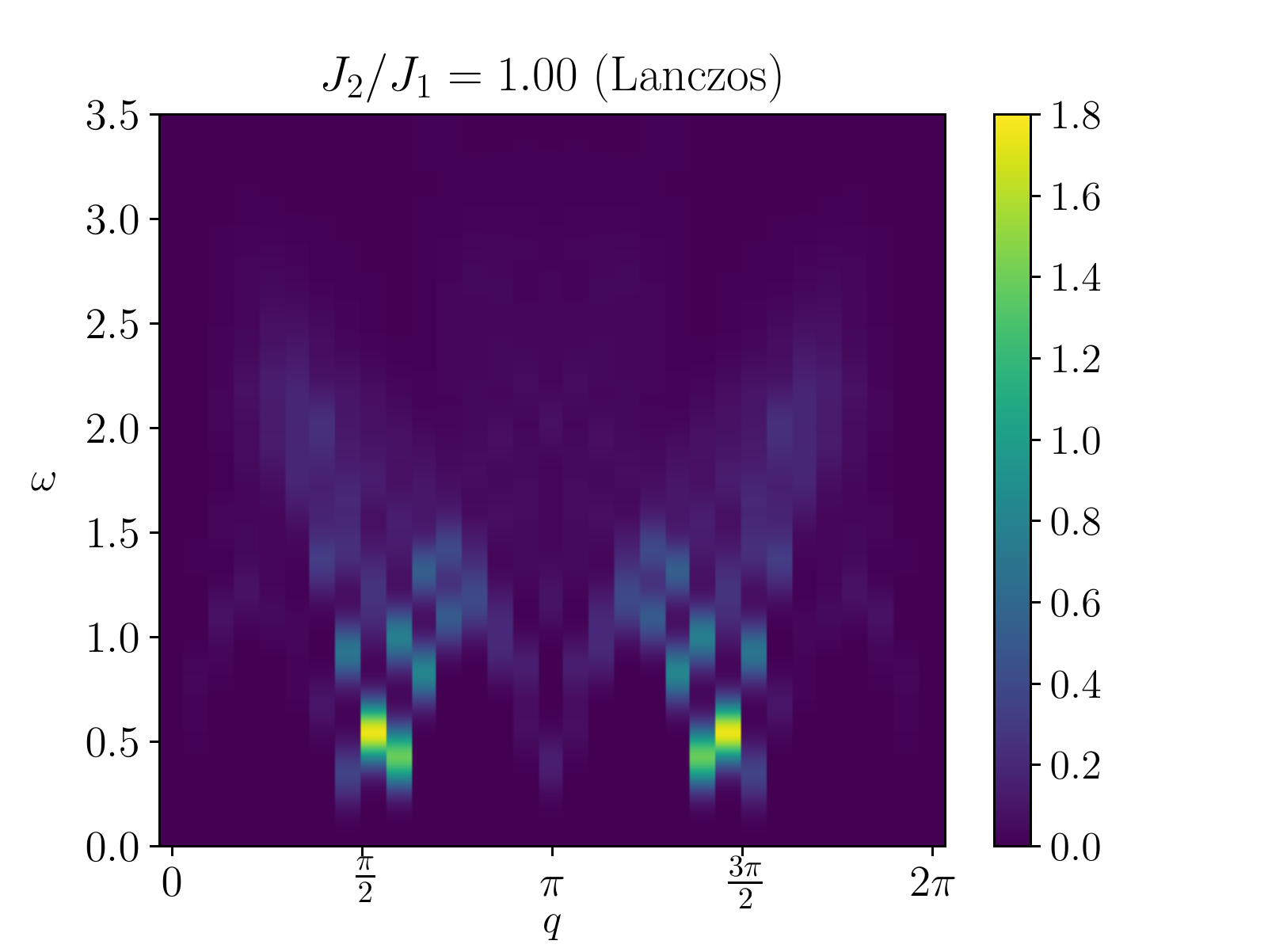}
\caption{\label{fig:L30}
Dynamical spin structure factor $S^{z}(q,\omega)$ for a chain of $L=30$ sites: comparison of variational and Lanczos results for 
different values of the frustrating ratio. The delta-functions in Eqs.~(\ref{eq:dsf}) and~(\ref{eq:Szz_practical}) have been replaced 
by normalized Gaussians with $\sigma=0.1 J_1$.}
\end{figure*}

\begin{figure*}
\includegraphics[width=0.4\textwidth]{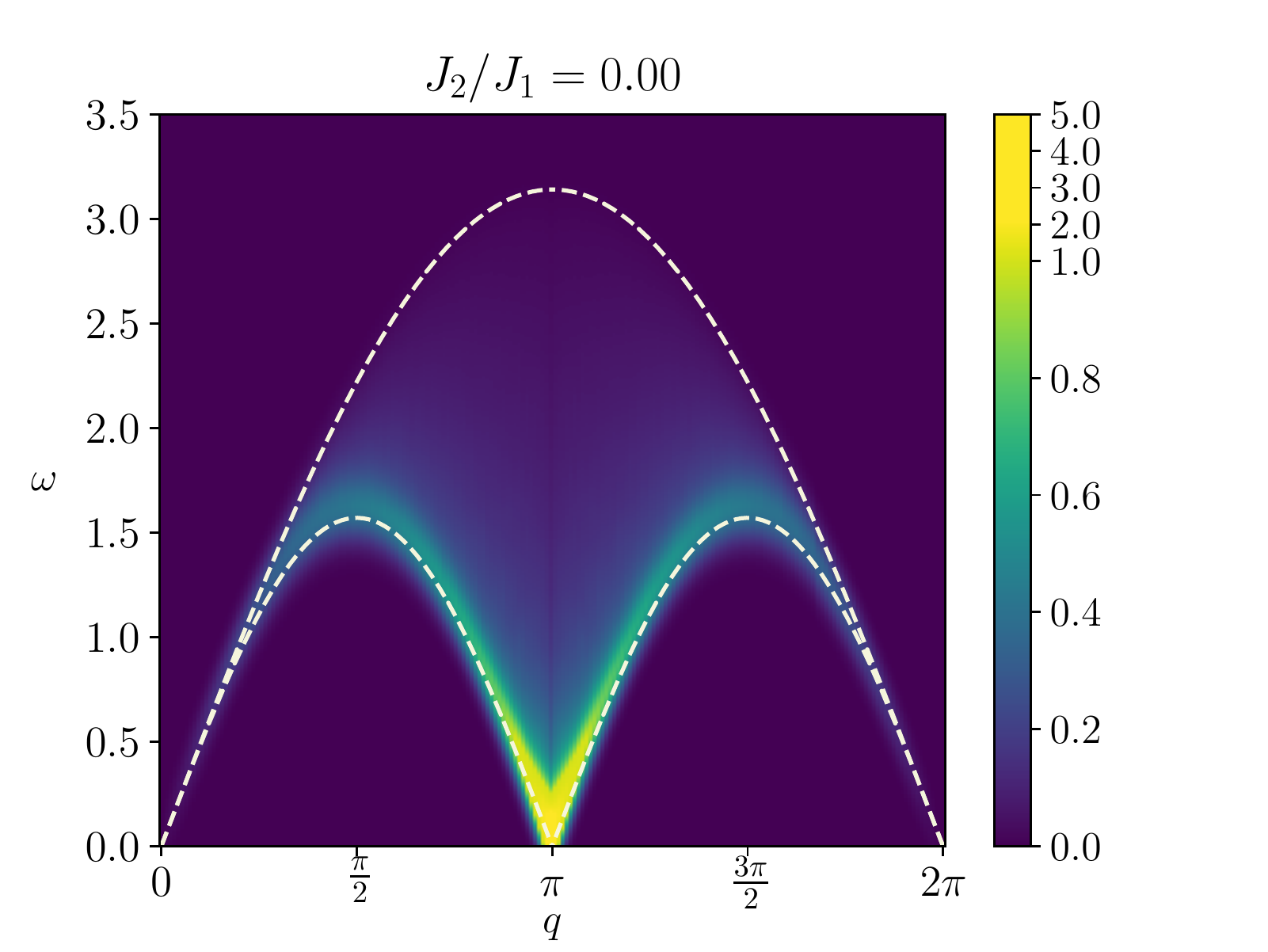}
\includegraphics[width=0.4\textwidth]{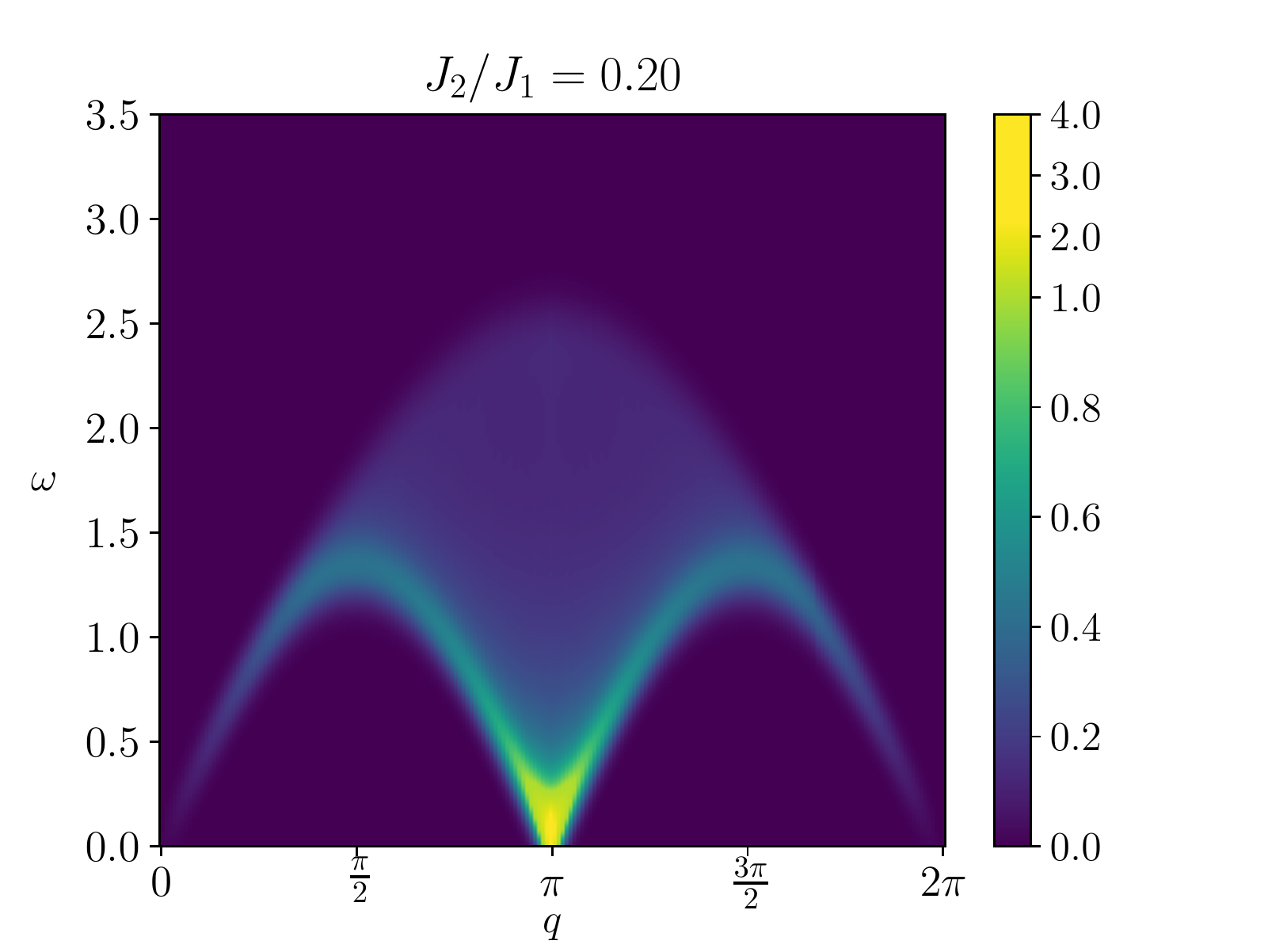}
\includegraphics[width=0.4\textwidth]{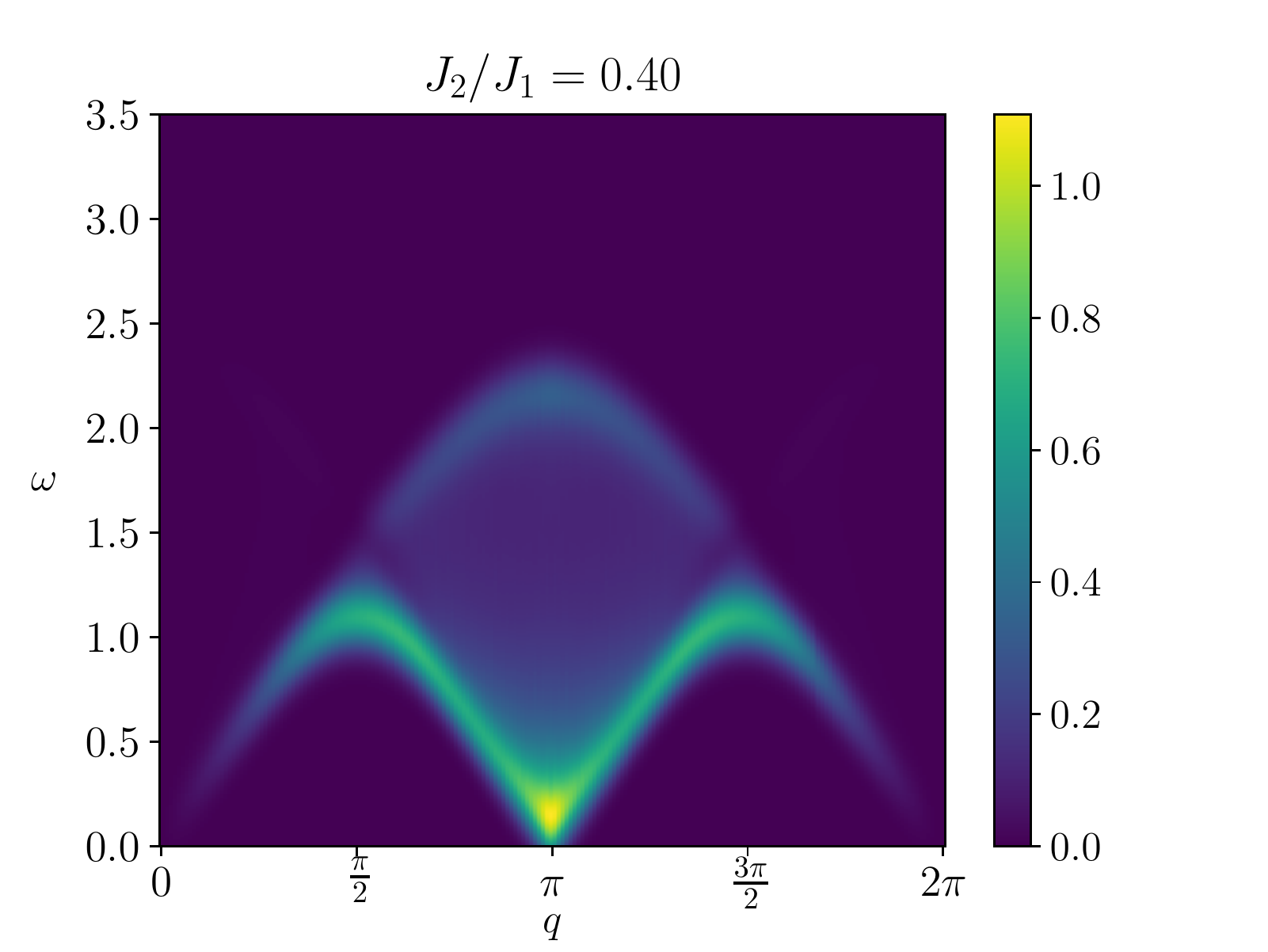}
\includegraphics[width=0.4\textwidth]{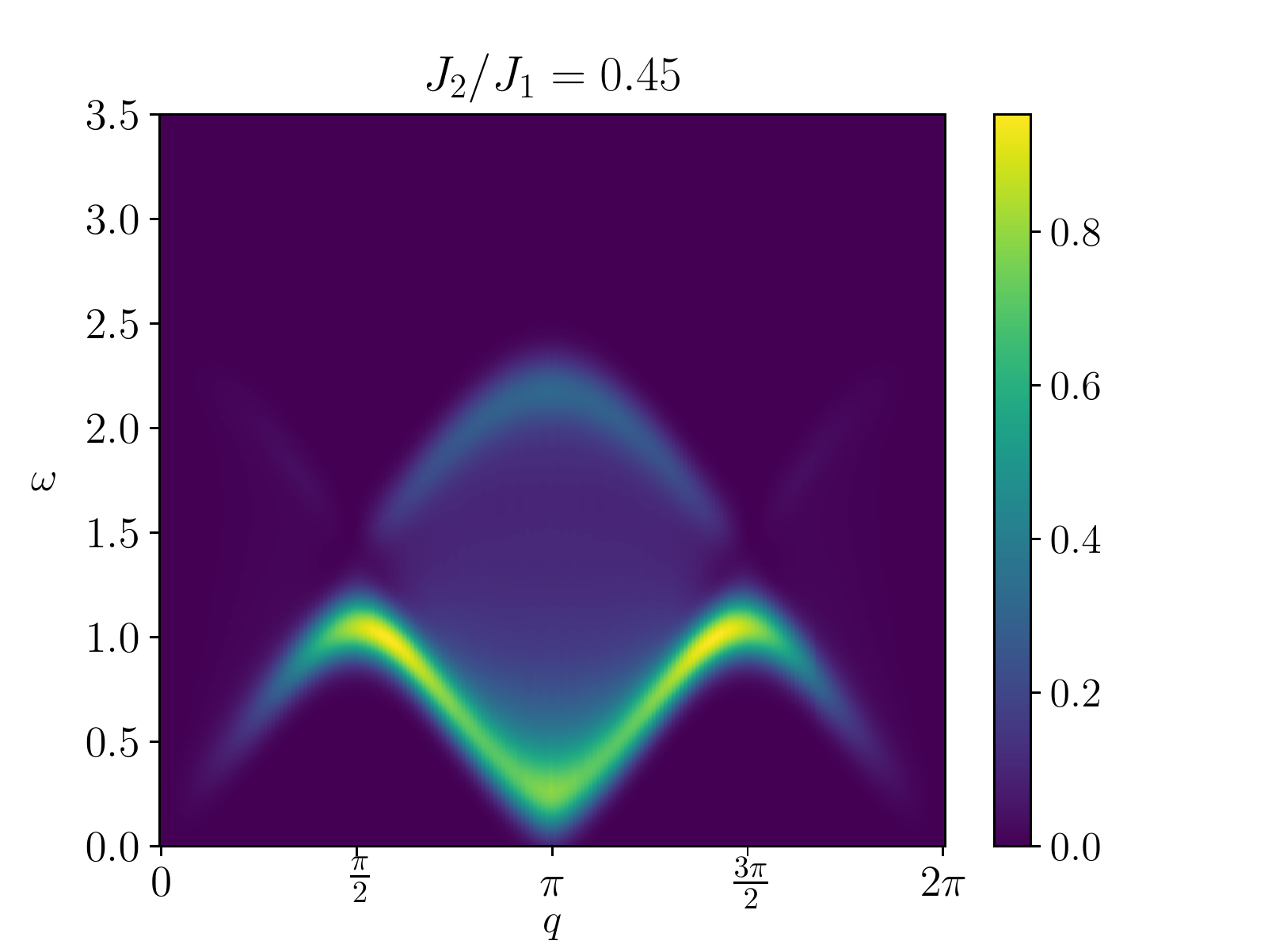}
\includegraphics[width=0.4\textwidth]{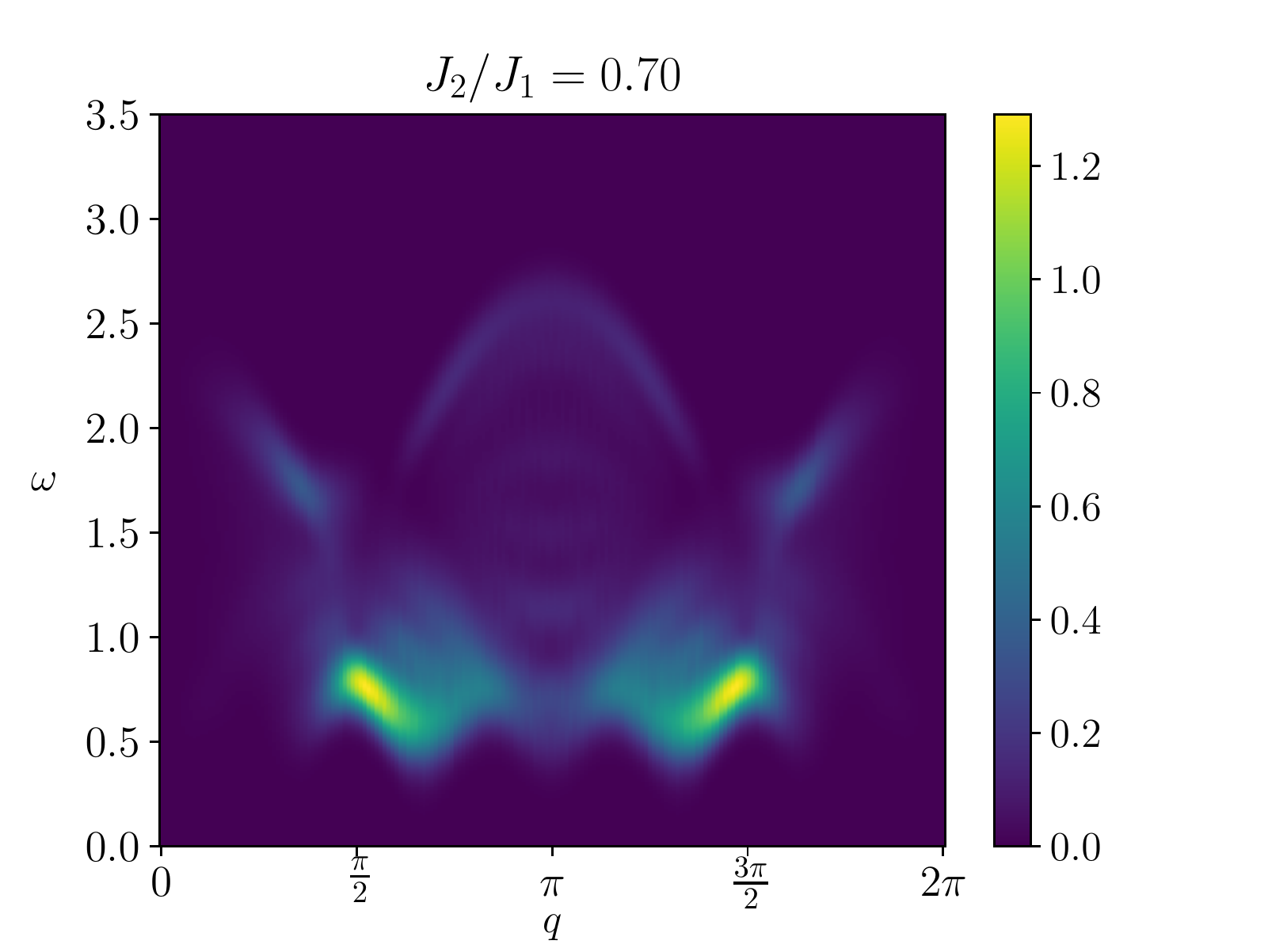}
\includegraphics[width=0.4\textwidth]{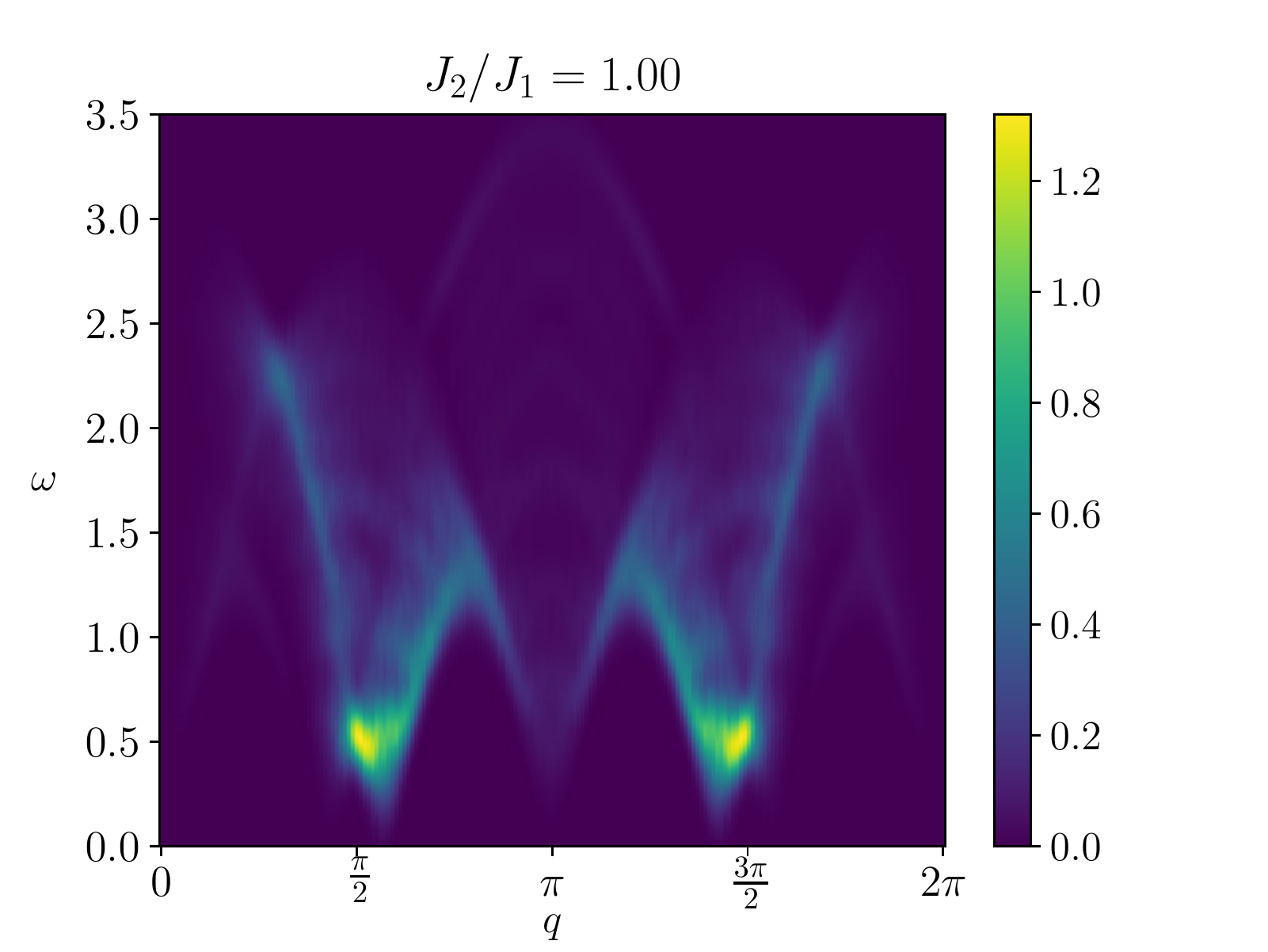}
\caption{\label{fig:L198}
Dynamical spin structure factor $S^{z}(q,\omega)$ for different values of the frustrating ratio and $L=198$ sites. The delta-functions 
in Eq.~(\ref{eq:Szz_practical}) have been replaced by normalized Gaussians with $\sigma=0.1 J_1$. The white dashed lines for $J_2=0$ 
indicate the lower and upper limits of the two-spinon contributions, Eqs.~(\ref{eq:spinon1}) and~(\ref{eq:spinon2}).}
\end{figure*}
%%%%%%%%%%%%%%%%%%%%%%%%%%%%%%%%%%%%%%%%%%%%%

\section{The ground-state variational wave function}\label{sec:wavefunction}

The phase diagram of the one-dimensional $J_1-J_2$ Heisenberg model is well-known~\cite{white1996}: for small values of the frustrating 
ratio, the system is gapless (i.e., a Luttinger fluid) with power-law spin-spin correlations, while for large values of $J_2/J_1$, 
the system is in a gapped phase characterized by long-range dimer order. In addition, (short-ranged) spin-spin correlations show an
incommensurate periodicity for $J_2/J_1 \gtrsim 0.5$. The critical (Kosterlitz-Thouless) point that separates gapless and gapped phases 
has been estimated with high accuracy, $(J_2/J_1)^c=0.241167 \pm 0.000005$~\cite{eggert1996}.

The simplest \textit{Ansatz} that can be used to describe both phases is the one obtained from a pure hopping Hamiltonian ${\cal{H}}_0$ 
with broken translational symmetry. This can be achieved by doubling the unit cell and taking different intra-cell ($t_1$) and inter-cell 
($t_1^\prime$) hoppings. When $t_1=t_1^\prime$, ${\cal{H}}_0$ recovers translational invariance and reduces to the case of free fermions 
in one-dimension, which have a Fermi sea ground state and gapless excitations. Instead, when $t_1^\prime \neq t_1$, there are two fermionic 
bands separated by a finite gap. The uniform and dimerized states are dubbed \textit{UFS} and \textit{DFS}, respectively. The accuracy 
for a cluster with $L=30$ sites is shown in the top panel of Fig.~\ref{fig:accuracyGS}. For $J_2/J_1 \lesssim 0.35$ the optimal wave 
function does not break the translational symmetry (i.e., $t_1^\prime=t_1$); by contrast, for larger values of the frustrating ratio, 
$t_1^\prime \ne t_1$. At the Majumdar-Ghosh point ($J_2/J_1=0.5$), one of the two hopping parameters is equal to zero, indicating that the 
wave function is a product of nearest-neighbor singlets. Here, the variational state becomes exact. Actually, the fully dimerized wave 
function remains the optimal solution for $J_2/J_1>0.5$, but its accuracy quickly worsens, since its energy is independent on $J_2$.

More accurate wave functions can be built from translationally invariant \textit{Ans\"atze}, which include both hopping and pairing terms 
(with $t_{R,R^\prime}=t_{|R-R^\prime|}$ and $\Delta_{R,R^\prime}=\Delta_{|R-R^\prime|}$). Nonetheless, even by considering translational
symmetry, a ``spontaneous symmetry breaking'' mechanism is possible after Gutzwiller projection is included, leading, for example, to dimer 
order~\cite{becca2011,kaneko2016}. Within a gapless regime, an extremely accurate state, dubbed \textit{WFA} is constructed from a 
fermionic Hamiltonian that contains first- and third-neighbor hoppings ($t_1$ and $t_3$), as well as first-neighbor pairing ($\Delta_1$). 
This choice gives a gapless fermionic band at $k=\pm \pi/2$ and can be stabilized up to $J_2/J_1 \approx 0.15$. For larger values of the 
frustrating ratio, the pairing term goes to zero and the wave function coincides with the \textit{UFS} state. A different possibility, 
which allows the existence of a gap in the fermionic spectrum, is given by taking first-neighbor hopping and both onsite ($\Delta_0$) 
and second-neighbor ($\Delta_2$) pairings. This \textit{Ansatz}, which is dubbed \textit{WFB}, is gapped unless $\Delta_0=-\Delta_2$. 
Optimizing the parameters of this wave function for $L=30$ sites, we find that it reduces to the simple \textit{UFS} state (i.e.,
$\Delta_0=\Delta_2=0$) for $J_2/J_1 \lesssim 0.1$. Then, the optimal pairing terms become non-zero and the wave function proves to be 
more accurate than the \textit{DFS} state and stable for all the values of the frustrating ratios (see Fig.~\ref{fig:accuracyGS}).

We expect that, in the thermodynamic limit, the gap in the fermionic spectrum will open in the vicinity of the exact transition point 
$(J_2/J_1)^c$ and will follow the behavior of the spin gap. However, it is extremely hard to locate this point by performing a finite 
size-scaling analysis, since the gap is exponentially small in an extended region after the critical point.

\section{Results}\label{sec:results}

Here, we present the numerical results for the spin structure factor of a one-dimensional chain. Let us start by considering a small
cluster with $L=30$ sites, where exact diagonalizations are possible by using the Lanczos method. First of all, we demonstrate that 
the variational results do not change appreciably when considering the wave function \textit{WFA} or \textit{WFB} to compute the 
dynamical structure factor (see Fig.~\ref{fig:vmc_lanczos}). Indeed, even though the latter state is about five times less accurate 
than the former one for $J_2=0$ (see Fig.~\ref{fig:accuracyGS}), the actual differences between the two dynamical calculations are 
negligible (and either option gives an excellent description of the exact results). Therefore, in the following, we consider only 
the \textit{WFB} wave function to compute the dynamical structure factor.

In order to best quantify the agreement between the variational and the exact calculations, we directly report $S^{z}(q,\omega)$ for 
several momenta $q$ as a function of the frequency $\omega$ for two values of the frustrating ratio, $J_2/J_1=0$ and $0.45$ (see 
Figs.~\ref{fig:lanczos1} and~\ref{fig:lanczos2}). Here, the delta-functions related to the exact and variational energies entering 
Eqs.~(\ref{eq:dsf}) and~(\ref{eq:Szz_practical}) have been replaced by normalized Gaussians with $\sigma=0.05 J_1$. The agreement is 
very good, not only for the unfrustrated case with $J_2=0$ (see Fig.~\ref{fig:lanczos1}), but also in the presence of a sizable frustration,
$J_2/J_1=0.45$ (see Fig.~\ref{fig:lanczos2}). Similar results are also obtained for larger values of the ratio $J_2/J_1$ (see below).
Therefore, it is expected that, within this approach, both gapless and gapped regimes are correctly described. The accuracy of the 
variational method is highlighted in Fig.~\ref{fig:accuracy_exc}, where we report the overlaps between the variational excited 
states of Eq.~(\ref{eq:var_excitations}) and the exact ones. In particular, whenever the excited states are well separated in energy, 
it is easy to match each exact excitation with a corresponding variational one (in this case, the overlap is very large, as for $J_2=0$). 
Instead, when two or even more excitations are close in energy, this correspondence is not easy to resolve and, for each variational 
state, we computed the overlap with all the exact states and plotted the maximum value. In this case, a reduction in the overlap is 
observed, as for a number of cases at $J_2/J_1=0.45$. Nevertheless, in all cases, the relevant excitations, which carry sizable 
spectral weigth, are well reproduced by the variational approach, and a reduced overlap is detected for states which do not contribute
much to the whole intensity of the dynamical structure factor.

Finally, the results obtained with the exact and variational approaches are compared in Fig.~\ref{fig:L30}, where $S^z(q,\omega)$ of 
a chain of $L=30$ sites is represented using color maps for $J_2/J_1=0.2$, $0.45$, $0.7$ and $1$. In all the cases, the variational 
results follow the exact ones, including the development of incommensurate features when $J_2/J_1>0.5$. In fact, by increasing the 
frustrating ratio, the intensity progressively shifts from $q=\pi$ (low energies) to $q=\pm \pi/2$ (high energies). At even larger 
values of $J_2/J_1$, the modes at $q=\pm \pi/2$ soften and eventually become gapless for $J_2 \to \infty$ (in this limit, the spin 
Hamiltonian consists of two decoupled Heisenberg models, one for each sublattice, and $J_2$ represents a nearest-neighbor superexchange 
on each sublattice). Remarkably, the variational approach is able to perfectly reproduce all the relevant features of the dynamical 
structure factor. Moreover, we find that the sum rule
\begin{equation}
\int d\omega S^z(q,\omega) = \langle \Psi_0| S^z_{-q} S^z_{q} |\Psi_0\rangle
\end{equation}
is satisfied within the errorbars for all the values of the frustrating ratio considered here. We mention the fact that the only case where 
our sampling technique fails is at the Majumdar-Ghosh point $J_2/J_1=0.5$, where the number of vanishing configurations in the ground-state 
wave function (exactly reproduced by our Gutzwiller-projected fermionic state) is exponentially large.

The results for a large cluster with $L=198$ sites are reported in Fig.~\ref{fig:L198} for $J_2/J_1=0$, $0.2$, $0.4$, $0.45$, $0.7$, and 
$1$.  In the unfrustrated case, it is known~\cite{karbach1997}, that most of the total intensity of the dynamical structure factor is 
carried by the two-spinon contributions. For these excitations the lower and upper energy limits are given by:
\begin{eqnarray}
\label{eq:spinon1}
\omega_{\rm lower} = \frac{\pi}{2} \left| \sin (q) \right|, \\
\omega_{\rm upper} = \pi \left| \sin \left (\frac{q}{2} \right ) \right|.
\label{eq:spinon2}
\end{eqnarray}
Indeed, we find that our dynamical structure factor is bounded by these limits and closely resembles the one that has been recently 
obtained with a Bethe \textit{Ansatz} approach~\cite{lake2013,caux2006}. 

It should be stressed that, for a relatively large region within the gapped phase, the value of the spin gap remains very small, since 
the transition from the gapless to the dimerized phase belongs to the Kosterlitz-Thouless universality class. Therefore, even for a relatively 
large system size, it is very hard to detect the presence of a finite gap in the excitation spectrum: for example, the dynamical structure 
factors at $J_2/J_1=0.2$ and $0.4$ (see Fig.~\ref{fig:L198}) look very similar, even though the former case corresponds to a gapless phase 
and the latter corresponds to a gapped spectrum. On this large cluster, the gradual shift of the intensity from $q=\pi$ to $q=\pm \pi/2$ is evident, 
as well as the presence of a ``rounding'' around $q=\pi$ within the gapped phase for $J_2/J_1<0.5$. Within such a large size, incommensurate 
features appear clearly for $J_2/J_1>0.5$; namely, the excitations with lowest energy move from $q=\pi$ to $q=\pm \pi/2$, giving rise to
a non-trivial form of the spectral function. These effects are determined by the gapped BCS spectrum, whose minima lie at incommensurate momenta. 
The rich structure of $S^z(q,\omega)$ is related to the fact that, in the limit $J_2/J_1 \to \infty$, the system decouples into two independent 
Heisenberg chains with coupling constant $J_2$. The Brillouin zone is then halved with respect to the case with $J_2=0$, and the dynamical 
structure factor is given by the repetition of the one of the pure Heisenberg model between $[0,\pi]$ and $[\pi,2\pi]$, scaled by $J_2/J_1$. 
For finite values of $J_2/J_1$ in the incommensurate phase, the spectral features at high energies are related to the lower and upper bounds 
of the two-spinon continuum that develops in the aforementioned limit.

\section{Conclusions}\label{sec:conclusions}

In conclusion, we have used a variational approach to study the dynamical spin structure factor of the frustrated $J_1-J_2$ model in one 
spatial dimension. Here, excitations at a given momentum $q$ are directly constructed from a Gutzwiller-projected fermionic wave function, 
thus avoiding any sign problem and/or analytic continuation from imaginary or real times to frequencies. In contrast to the original definition
where these excitations have $S^{z}=1$~\cite{li2010}, here we have considered states with $S^{z}=0$, which allow us to have a much simpler 
Monte Carlo sampling. Indeed, within our technique, the dynamical structure factor $S^{z}(q,\omega)$ can be computed for all momenta $q$ 
within a {\it single} Monte Carlo simulation.
 
We have reported the unprecedented accuracy of this method, not only at or close to the integrable point with $J_2=0$, but also for generic 
values of the frustrating ratio. The remarkable advantage of this variational procedure is given by the fact that the relevant part of the 
low-energy spectrum can be described by considering particle-hole excitations on top of a fixed ``reference'' state. Indeed, once the 
variational wave function has been optimized for the ground state, only $O(L)$ parameters for each $q$ are used to reproduce the low-energy 
part of the spectrum. This fact suggests that Gutzwiller-projected fermionic wave functions not only may accurately reproduce the ground-state
properties of frustrated spin models, but also constitute a good framework to generate low-energy excitations. 

This work shows that the present variational approach to compute the dynamical structure factor is very promising, especially in the case of 
two-dimensional frustrated spin models, for which a straightforward generalization is possible. The reliability of the results for different 
(gapped and gapless) phases and the possibility to consider relatively large sizes make this method very suitable not only for theoretical 
investigations of other spin models, but also for direct comparisons with neutron-scattering experiments.

\bibliographystyle{apsrev4-1}

\begin{thebibliography}{99}
\bibitem{note1} In many cases, the set of states $\{|q,R\rangle \}$ is linearly dependent. 
                This reduces the effective number of excitations $|\Psi_n^q\rangle$ which can be constructed.
\end{thebibliography}

\end{document}